\documentclass[12pt]{article}
\usepackage{amssymb, amsmath, bm, float, epsfig, color}
\usepackage[centering]{geometry}
\usepackage[utf8]{inputenc}
\definecolor{mygrn}{rgb}{0,0.5,0}

\begin{document}
\title{
\vspace{-30pt}
\begin{flushright}
\normalsize DESY 18-022 \\
KIAS-P18014 \\
WU-HEP-18-03 \\*[55pt]
\end{flushright}
{\Large \bf Seesaw mechanism in magnetic compactifications
\\*[20pt]}
}

\author{%XXX~XXX,$^X$\footnote{E-mail address: ???} \, and \,
Makoto~Ishida,$^1$\footnote{E-mail address: ncanis3@fuji.waseda.jp} \quad
Kenji~Nishiwaki,$^2$\footnote{E-mail address: nishiken@kias.re.kr} \, and \, 
Yoshiyuki~Tatsuta,$^{1, \, 3}$\footnote{E-mail address: yoshiyuki.tatsuta@desy.de}\\*[30pt]
$^1${\it \normalsize Department of Physics, Waseda University, Tokyo 169-8555, Japan}\\
$^2${\it \normalsize School of Physics, Korea Institute for Advanced Study~(KIAS), Seoul 02455, Republic of Korea}\\
$^3${\it \normalsize Deutsches Elektronen-Synchrotron DESY, Hamburg 22607, Germany}\\*[55pt]
}

%0
\date{
\centerline{\small \bf Abstract}
\begin{minipage}{0.9\textwidth}
\medskip\medskip 
\small
{In this paper, we explore a new avenue to a natural explanation of the observed tiny neutrino masses with a dynamical realization of the three-generation structure in the neutrino sector.
Under the magnetized background based on $T^2/Z_2$, matter consists of multiply-degenerated zero modes and the whole intergenerational structure is dynamically determined.
In this sense, we can conclude that our scenario is favored by minimality, where no degree of freedom remains to deform the intergenerational structure by hand freely.
Under the consideration of brane-localized Majorana-type mass terms for {an} $SU(2)_L$ singlet neutrino, it is sufficient to introduce one Higgs doublet {for reproducing the observed neutrino data}.
In all reasonable flux configurations with three right-handed neutrinos, phenomenologically acceptable parameter configurations are found.}
\end{minipage}}

\begin{titlepage}
\maketitle
\thispagestyle{empty}
%\clearpage
%\tableofcontents
%\thispagestyle{empty}
\end{titlepage}
\tableofcontents

%0
\section{Introduction}
The standard model (SM) of elementary particles has been completed by the discovery of the last puzzle piece, i.e., the Higgs boson in 2012 \cite{Aad:2012tfa, Chatrchyan:2012ufa}.
Before discovering the Higgs boson, the non-vanishing neutrino masses reported in 1996 {have demanded} that the SM must be extended to its neutrino sector with right-handed neutrinos.
Recently, the neutrino flavor structures have been steadily revealed from the viewpoints of neutrino oscillations \cite{Abe:2013hdq, Adamson:2016tbq} and cosmological behaviors of neutrinos~{\cite{Patrignani:2016xqp}}.
The precise theoretical investigation in the neutrino flavor structure {can be} one of the main pillars in the modern particle physics.

In contrast to the other SM three-generation fermions, i.e., quarks and charged leptons, several experiments have shown that the neutrinos have tiny masses around eV scale.
This experimental result {implicitly tells} that there {may be} a particular mechanism only in the neutrino sector.
One of the mechanisms which can explain the tiny neutrino masses is the (type I) seesaw mechanism {\cite{Minkowski:1977sc,Yanagida:1979as,GellMann:1980vs,Mohapatra:1979ia,Schechter:1980gr}} by means of the right-handed neutrino Majorana mass term.
Only by adding the heavy Majorana mass term at some high scale in addition to the Dirac mass term, the effective neutrino masses can be small enough for the experimental results, even if the Dirac mass term appears around the electroweak (EW) scale.
Although the seesaw {mechanism} is quite simple and {beneficial} in many scenarios, there is still an ambiguous point in the detailed structures of the Dirac and Majorana mass matrices.
In usual bottom-up approaches where one assumes some extensions to the SM, it is generically difficult to theoretically determine the concrete entries and values in the mass matrices.
Then, in order to control the matrix entries, one {pursuits} the models with the continuous flavor symmetry \cite{Froggatt:1978nt}, the discrete flavor symmetry \cite{Ishimori:2010au} and the extra dimension(s) \cite{ArkaniHamed:1999dc}, for instance.

As well as the matrix entries in the neutrino sector, to understand an origin of the three-generation structure behind the SM fermions is still a challenging issue.
Among recent topics, an interesting attempt to reveal the generation structure is to add magnetic fluxes on compactified extra dimensions.
In particular, the magnetic fluxes turned on the torus provide multiple massless wavefunctions after the Kaluza--Klein (KK) decomposition of fields \cite{Bachas:1995ik, Cremades:2004wa}.
The same happens in the extensions to toroidal orbifolds {\cite{Braun:2006se, Abe:2008fi, Fujimoto:2013xha}} in particular those with discrete Wilson line phases {\cite{Abe:2013bca, Abe:2014noa, Buchmuller:2015eya, Matsumoto:2016okl, Kobayashi:2017dyu}}.
Since such multiple massless modes belong to the same representation of fields, the multiplicity of massless modes should be identified as the generation structure in the SM.
Many model constructions by means of the mechanism have been done especially in the past five years, for example, supersymmetric models \cite{Abe:2012fj, Abe:2016jsb}, non-supersymmetric models \cite{Abe:2014vza}, systematic searches of three-generation models {\cite{Abe:2013bba, Abe:2015yva, Fujimoto:2016zjs}}, three-generation models with broken supersymmetry \cite{Buchmuller:2015jna}, quark and charged lepton mass matrices from bulk overlap integrals \cite{Abe:2014vza, Abe:2016jsb} and brane-localized Yukawa couplings \cite{Buchmuller:2015jna, Buchmuller:2017vho, Buchmuller:2017vut}, mass spectra in the presence of brane-localized {mass} terms \cite{Ishida:2017avx} and cosmological inflation model \cite{Higaki:2016ydn}, applications to volume moduli stabilization \cite{Buchmuller:2016dai,Buchmuller:2016bgt}.

{It is noted that we face difficulties in generating} the neutrino Majorana mass term in the previous model buildings based on extra dimensional fluxes \cite{Abe:2012fj}.\footnote{If the flux compactification is derived from superstring theory, the Majorana mass term can be induced by D-brane instanton effects \cite{Ibanez:2006da, Ibanez:2012zz,Hamada:2014hpa,Kobayashi:2015siy}.}
In this paper, the brane-localized mass term(s) on a toroidal orbifold $T^2/Z_2$ analyzed in \cite{Ishida:2017avx} is applied to the type I seesaw scenario.
Then, the Majorana mass matrix is analytically given by the localized neutrino masses and concrete values are determined by the values of zero-mode wavefunctions evaluated at orbifold fixed points.
In addition, the Dirac mass matrix originates from the Yukawa couplings which are analytically calculated by overlap integrals of zero-mode wavefunctions.
Thus, the seesaw mechanism in terms of brane-localized neutrino masses at fixed points of $T^2/Z_2$ with extra dimensional fluxes is a simple and typical scenario.

This paper is organized as follows.
In Sec.\,2, we briefly review several ingredients for the seesaw mechanism in the orbifold $T^2/Z_2$ on flux background.
In Sec.\,3, {five} patterns of neutrino mass matrices realized on such an orbifold are numerically analyzed and compared with observed values by recent neutrino oscillation experiments.
We make conclusion in Sec.\, 4.

\section{Review of $T^2/Z_2$ orbifold with fluxes}
In this section, we briefly review the six-dimensional (6D) compactification on {the} orbifold $T^2/Z_2$ with fluxes, and {show the} KK mass spectra and wavefunctions as well as (three-point) Yukawa coupling constants.
In seesaw {scenarios}, such Yukawa couplings between neutrinos and the Higgs boson provide the Dirac mass matrix after the Higgs boson develops its vacuum expectation value (VEV).
On another hand, the Majorana mass term for the seesaw originates from the existence of brane-localized term(s) at orbifold fixed points of $T^2/Z_2$.
This section is mainly based on \cite{Ishida:2017avx, Cremades:2004wa, Abe:2008fi, Abe:2008sx}.

\subsection{Flux background and Yukawa couplings}
We consider the 6D gauge theory compactified on $\mathcal{M}^4\times T^2/Z_2$.
Here, $\mathcal{M}^4$ is the {four-dimensional}~{(4D)} Minkowski spacetime and we choose $T^2/Z_2$ to be extra dimensions of our model.

\begin{figure}[t]
\centering
\includegraphics[width=0.4\textwidth]{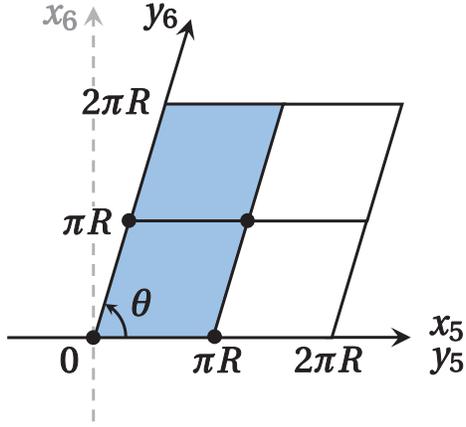}
\caption{{The fundamental domain of the orbifold $T^2/Z_2$ {is shown}.
{The} shaded region is a fundamental region and {the} black dots {represent the} four fixed points of the orbifold.
This figure is drawn for a generic complex structure $\tau \in \mathbb{C}$, where an angle between $y_5$ and $y_6$ is given as $\cos \theta \equiv {\rm Re} \, \tau/|\tau|$.}}
\label{fps}
\end{figure}

We first start {in} a two-dimensional torus $T^2$.
{We} define two {oblique} coordinates $y_5$ and $y_6$ as coordinates of $T^2$ and {such coordinates} are often conveniently expressed by {the} complex coordinate $z=(y_5+\tau y_6)/(2\pi R)$, with a complex structure modulus $\tau\in\mathbb{C} \,\, {({\rm Im}\, \tau>0)}$ and a radius $R$, {where a schematic picture is depicted in Fig.~\ref{fps}}.
{Notice that the radius $R$ is associated with a compactification scale $M_C \sim 1/R$.}
{The} toroidal orbifold $T^2/Z_2$ is obtained by {the identifications in the} two-dimensional extra dimensions under the toroidal periodicities and the $Z_2$ rotation,
\begin{eqnarray}
z\sim z+1\sim z+\tau\sim-z.
\end{eqnarray}
In accordance with the above identifications, there {appear} four fixed points on $T^2/Z_2$, i.e., {at $z =$} $0, 1/2, \tau/2$ and $ (1+\tau)/2$, as described in Fig.\,\ref{fps}. 

Kinetic terms of 6D {Weyl} fermions and scalars are given as
\begin{eqnarray}
\label{Lkin}
\mathcal{L}_{\rm kin}=\int d^4x\int_{T^2}d^2z \, {\left\{ i\bar{\Psi}\Gamma^MD_M\Psi  + (D_M\Phi)^\dagger(D^M\Phi)\right\}}.
\end{eqnarray}
In Eq.\,(\ref{Lkin}), $M$ runs over $0,1,2,3, 5,6$, $\Gamma^M$ denotes {the gamma} matrices describing the {Clifford} algebra in six dimensions, and $D_M=\partial_M-iqA_M$ denotes a covariant derivative under {a} $U(1)$ gauge symmetry.
In the following, we discuss the toroidal case at the first step and, then extend it to the toroidal orbifold case.
In the six-dimensional action, we assume that the vector potential $A_{m} \,\, (m=5,6)$ possesses classical non-trivial background $b=\int_{T^2}F$ of the field strength $F=(ib/2{\rm Im}\tau)dz\wedge d\bar{z}$:
\begin{eqnarray}
A^{(b)}(z)=\frac{b}{2{\rm Im}\,\tau}{\rm Im}\,(\bar{z}dz).
\label{connection}
\end{eqnarray}
The consistency condition provides the quantization condition of fluxes:
\begin{eqnarray}
\frac{qb}{2\pi}=M\in\mathbb{Z}.
\label{Dirac}
\end{eqnarray}
It should be mentioned that there is a controversial point about the Dirac charge quantization condition \eqref{Dirac}.
As naturally expected, several papers \cite{Bachas:1995ik, Braun:2006se, Buchmuller:2015eya} claim that a flux density of bulk constant flux is twice as that on the original torus.
On the other hand, some of research groups \cite{Abe:2008fi, Abe:2013bca} have investigated it in the framework of conformal field theory and have reported different quantization conditions, i.e., the same one as the original torus.
One of their claims is that Eq.\,\eqref{connection} behaves as an appropriate $U(1)$ connection on the orbifold even with the original charge quantization \eqref{Dirac}.
Throughout this paper, we follow the latter condition.

We perform the KK decomposition of six-dimensional fields.
The six-dimensional {Weyl fermions and scalars} are decomposed as
\begin{eqnarray}
\Psi(x^\mu,z)=\sum_n\chi_n(x^\mu)\otimes\psi_n(z),\\
\Phi(x^\mu,z)=\sum_n\varphi_n(x^\mu)\otimes\phi_n(z).
\end{eqnarray}
The fermion wave functions in {the} extra two directions are determined as eigenstates of the Dirac operator in extra dimensional directions,
\begin{eqnarray}
i\Gamma^mD_m\psi_n(z)=m_n\psi_n(z),
\end{eqnarray}
where $m=5, 6$ and $m_n\  (n=0,1,2\cdots)$ denote the KK mass {spectrum}.
Hence, zero-mode equations for $n=0$ {($m_0 = 0$)} are given {in terms of $D \equiv D_5+\tau D_6$} as
\begin{eqnarray}
\label{zeromodeeq}
D \, \psi_+(z)=0,\ \qquad D^\dagger \, \psi_-(z)=0,
\end{eqnarray}
where the two-dimensional spinor is decomposed into $\psi_0=\left(\psi_+,\psi_-\right)^T$ with {the two-dimensional internal} chiralities.
In several appropriate boundary conditions associated with necessary gauge transformations, we describe zero-mode wave functions analytically in terms of the Jacobi theta function \cite{Cremades:2004wa},
\begin{align}
\psi_+^{j}(z) &\equiv \Theta^{j, M}(z) \notag\\
&= {\mathcal{N}_M} \, e^{i\pi Mz {\rm Im}\, z /{\rm Im}\, \tau} \cdot \vartheta \label{functionform}
\begin{bmatrix}
j/M\\[5pt]
0 
\end{bmatrix}
\left(Mz,M\tau\right) \qquad (j=0, 1,2, \cdots, M-1),
\end{align}
for $M>0$.
Here, ${{\cal N}_M} = (2M/{\cal A}^2)^{1/4}$ is a normalization constant \cite{Cremades:2004wa}, {where $\mathcal{A}$ represents the area of the torus and ${\cal N}_M$ has mass dimension $+1$}.
%(For $M<0$, the zero-mode wavefunction is complex conjugate of that for $M>0$, as discussed in \cite{Cremades:2004wa}.)
If {the} flux number $M$ is positive (negative), there is no normalizable solution {in} $\psi_-$ ($\psi_+$).
Notice that Eq.\,\eqref{functionform} tells that Eq.\,(\ref{zeromodeeq}) has {$|M|$-independent} solutions.
Hence we can identify this degeneracy of zero-modes with a family structure for particles in the four-dimensional effective theory.
Therefore, if we introduce a non-zero magnetic flux, a chiral structure of the Weyl spinor appears in the four-dimensional effective theory.
%The same family-generating mechanism holds for the Majorana spinor.
%For both Weyl and Majorana spinors,
{The form of the KK masses is} obtained as
\begin{gather}
m_n^2 = \frac{4\pi M}{\cal A}n.
\end{gather}
{Also, we} similarly compute a scalar wave function {as} the same as the fermionic one in Eq.\,(\ref{functionform}).
{There} is no massless (zero-)mode in the scalar field.
In other words, the lowest KK mass is non-vanishing, {where the KK mass spectrum for scalars is} given as
\begin{gather}
m_n^2 = \frac{4\pi M}{\cal A} \left(n+\frac12 \right).
\label{KKscalar}
\end{gather}
	
%As mentioned above, we treat the $T^2/Z_2$ orbifold in this paper.
{Next}, we move to $Z_2$ eigenstates of zero-modes under the twisted $Z_2$ projection, i.e., $z \sim -z$.
{The} eigenstates of zero-modes on $T^2/Z_2$ are expressed as linear combinations of {those in $T^2$ shown in} Eq.\,(\ref{functionform}) \cite{Braun:2006se, Abe:2008fi},
\begin{eqnarray}
\Theta^{j,M}_{T^2/Z_2,\eta}(z)&=&\frac{1}{\sqrt{2}}\left(\Theta^{j,M}(z)+\eta\Theta^{j,M}(-z)\right)\nonumber\\
\label{Z2wavefunc}
&=&\frac{1}{\sqrt{2}}\left(\Theta^{j,M}(z)+\eta\Theta^{M-j,M}(z)\right),
\end{eqnarray}
where $\eta$ denotes {a $Z_2$ parity} so as to be the $Z_2$ even (odd) {as} $\eta=+1\ (-1)$.\footnote{
{In addition to the $1/\sqrt{2}$ factor, we should reshape the normalization factor as $(2 M/\mathcal{A}^2)^{1/4}/\sqrt{1 + \delta_{j, M/2}}$ (from $(2 M/\mathcal{A}^2)^{1/4}$) {when we take the range $T^2$ in the $d^2 z$ integration}~\cite{Ishida:2017avx}.}
}
As calculated in \cite{Abe:2008fi, Abe:2013bca}, the relation between flux numbers and the numbers of zero-mode wave functions are obtained as Tab.\,\ref{T2Z2generation}.

\begin{table}[t]
\begin{center}
\begin{tabular}{|c|cccccccccc|cc|}\hline
$M$ & $0$ & $1$ & $2$ & $3$ & $4$ & $5$ & $6$ & $7$ & $8$ & $\cdots$ & $2k$ & $2k+1$\\ \hline
$\eta=+1$ & $1$ & $1$ & $2$ & $2$ & $3$ & $3$ & $4$ & $4$ & $5$ & $\cdots$ & $k+1$ & $k+1$\\
$\eta=-1$ & $0$ & $0$ & $0$ & $1$ & $1$ & $2$ & $2$ & $3$ & $3$ & $\cdots$ & $k-1$ & $k$ \\ \hline
\end{tabular}
\caption{The relation between flux numbers and the numbers of zero-mode wave functions.}
\label{T2Z2generation}
\end{center}
\end{table}

Using the above analytic form of zero-mode eigenstates on $T^2/Z_2$, we can also analytically calculate Yukawa couplings as an {overlap} integral of three zero-modes.
First, we show {the form of} Yukawa couplings on $T^2$, and then extend them to those of $T^2/Z_2$.
{Among only zero-modes, effective} Yukawa couplings {after dimensional reduction} can be computed from the six-dimensional Lagrangian,
\begin{eqnarray}
\mathcal{L}_{\rm Yukawa} &=&\int_{T^2}d^2z\left\{-g\overline{\Psi_1} \Psi_2\Phi+{\rm h.c.}\right\} \nonumber\\
&{\supset}&-\left(g\int_{T^2}d^2z \, \Theta^{i,M_1}(z)\Theta^{j,M_2}(z) \bigl( \Theta^{k,M_3}(z) \bigr)^*\right)\overline{\chi_1^{{i}}}{\chi_2^{{j}}}\varphi^k+{\rm h.c.},
\end{eqnarray}
{where we suppose $|M_3| \geq |M_1|, |M_2|$.}
{It is noted that the coefficient $g$ has mass dimension $-1$.}
Hence, Yukawa couplings can be expressed as
\begin{eqnarray}
\label{yukawa1}
Y^{ijk}=g\int_{T^2}d^2z \, \Theta^{i,M_1}(z)\Theta^{j,M_2}(z) \bigl( \Theta^{k,M_3}(z) \bigr)^*.
\end{eqnarray}
%Furthermore, the consistency condition on magnetic fluxes requires $|M_1|\pm|M_2|=|M_3|$. Using the formula;
%\begin{eqnarray}
%\Theta^{i,M_1}\Theta^{i,M_2}=\sum_{m\in\mathbb{Z}_{M_3}}\Theta^{i+j+M_1m,M_3}\times\vartheta\left[
%\begin{array}{c}
%\frac{M_2I-M_1j+M_1M_2m}{M_1M_2M_3}\\
%0
%\end{array}
%\right]\left(0,\tau M_1M_2M_3\right),
%\end{eqnarray}
%and the orthonormal condition: $\int_{T^2}d^2z\Theta^{j,M}\Theta^{k,M}=\delta_{jk}$, we can compute Eq.~(\ref{yukawa1}) and express analytically as follows;
As calculated in \cite{Cremades:2004wa}, it is straightforward to perform an integration in Yukawa couplings and we finally obtain
\begin{eqnarray}
\label{T2yukawa}
Y^{ijk}=g \,  {\frac{ \mathcal{N}_{|M_1|} \mathcal{N}_{|M_2|} }{ \mathcal{N}_{|M_3|}}} \, \sum_{m=0}^{|M_3|-1}\vartheta
\begin{bmatrix}
\frac{M_2i -M_1j+M_1M_2m}{M_1M_2M_3}\\[5pt]
0 
\end{bmatrix}
(0,\tau M_1M_2M_3)\times\delta_{i+j+M_1m,k+M_3l},
\end{eqnarray}
{where the overall coupling is a dimensionless factor.}

Now, we extend this formula of Yukawa couplings to those of $T^2/Z_2$.
In this case, Yukawa couplings can be expressed as
\begin{eqnarray}
Y^{ijk}_{T^2/Z_2}=g\int_{T^2}d^2z \, \Theta^{i,M_1}_{T^2/Z_2,\eta_1}(z) \Theta^{i,M_2}_{T^2/Z_2,\eta_2} (z) \bigl(\Theta^{i,M_3}_{T^2/Z_2,\eta_3}(z) \bigr)^*.
\end{eqnarray}
Using Eqs.\,(\ref{Z2wavefunc}) and {(\ref{yukawa1})}, Yukawa couplings on $T^2/Z_2$ are described as
\begin{eqnarray}
Y^{ijk}_{T^2/Z_2}&=&\frac{1}{2\sqrt{2}}\left(Y^{ijk}+\eta_1Y^{(M_1-i)jk}+\eta_2Y^{i(M_2-j)k}+\eta_3Y^{ij(M_3-k)}\right.\nonumber\\
&\ &\ \ \ \ \ \ \ \ +\eta_1\eta_2Y^{(M_1-i)(M_2-j)k}+\eta_2\eta_3Y^{i(M_2-j)(M_3-k)}+\eta_1\eta_3Y^{(M_1-i)j(M_3-k)}\nonumber\\
&\ &\ \ \ \ \ \ \ \ \left.+\eta_1\eta_2\eta_3Y^{(M_1-i)(M_2-j)(M_3-k)}\right).
\label{yukawas}
\end{eqnarray}
The concrete entries in Yukawa couplings are analytically written in the {appendices} of \cite{Abe:2015yva, Abe:2008sx} for arbitrary configurations of fluxes and (discrete) Wilson lines.
{A selection rule is found} that Yukawa interaction terms in the Lagrangian must be invariant under the $Z_2$ parity transformation.
Thus, we have to only consider the four patterns for $\eta_1, \eta_2, {\rm and}\ \eta_3$ as
{$(\eta_1, \eta_2, \eta_3) = (+1, +1, +1),\, (+1, -1, -1), \, (-1, +1, -1), \, (-1, -1, +1)$}.

%\begin{table}[t]
%\begin{center}
%\begin{tabular}{|ccc|}\hline
%$\eta_1$ & $\eta_2$ & $\eta_3$\\ \hline
%$+1$ & $+1$ & $+1$\\
%$+1$ & $-1$ & $-1$\\
%$-1$ & $+1$ & $-1$\\
%$-1$ & $-1$ & $+1$\\ \hline
%\end{tabular}
%\caption{The possible patterns for $\eta_1, \eta_2$ and $\eta_3$.}
%\label{etapattern}
%\end{center}
%\end{table}

\subsection{Brane-localized Majorana mass terms}
In this paper, we {focus on} {the Majorana mass terms localized at the} orbifold fixed points in \cite{Ishida:2017avx},
\begin{eqnarray}
\mathcal{L}_{\rm brane} = - {\frac{1}{2}} \int_{{T^2}}d^2z \, \sum_{k=1}^4h_k  \overline{(\Psi_R)^C} \Psi_R \, \delta^2(z-z_k) + \text{h.c.},
\label{axion}
\end{eqnarray}
where $h_k \,\, (k=1,2,3,4)$ are constants {with mass dimension $-1$}.
Here, we assume that only the component with the 4D right-hand chirality {$(R \,)$} of the 6D Weyl spinor $\Psi$ contributes to the terms.\footnote{As explicitly discussed in~\cite{Frere:2010ah}, one 6D Weyl spinor is not sufficient for constructing 6D Majorana-type mass terms.
See also~\cite{Pilaftsis:1999jk,Dudas:2005vn}.}
In addition, the {4D Weyl} field $\Psi_R$ carries a $U(1)$ charge (or flux). 
Unless the $U(1)$ symmetry generating family structures is broken, the above localized Majorana mass term cannot be written down.
{Constructing a concrete model by introducing a scalar field for spontaneous breaking of the $U(1)$ (or embedding our setup into more fundamental theories) is beyond the interest of our paper on the observed neutrino profiles}.\footnote{Note that in a supergravity extension of our model an axion appears \cite{Buchmuller:2015eya} which can be used to make the mass term \eqref{axion} invariant under the $U(1)$ symmetry \cite{Buchmuller:2017vut}.}
For the moment, we assume an appropriate $U(1)$ breaking mechanism.
In the final section, we will comment on how to treat the $U(1)$ symmetry {in details}.
The superscript $C$ denotes the {4D} charge conjugation and $z_k \,\, (k=1,2,3,4)$ denote the $Z_2$ fixed points, i.e., 
\begin{gather}
z_1 = 0, \qquad z_2 = 1/2, \qquad z_3=\tau/2, \qquad z_4=(1+\tau)/2{.}
\end{gather}
The effective Majorana mass matrix in the low energy effective Lagrangian can be computed as
%%%%%
\begin{eqnarray}
\mathcal{L}_{\rm brane} &=&- {\frac{1}{2}} \sum_{k=1}^4h_k \left(\int_{{T^2}}d^2z \, \Theta^{i,M}_{T^2/Z_2,\eta}(z)\Theta^{j,M}_{T^2/Z_2,\eta}(z)\delta^2(z-z_k)\right) {\overline{(\chi^i_R)^{C}} \chi^j_R}  + \text{h.c.} \nonumber \\
&=& - {\frac{1}{2}} \left( \sum_{k=1}^4h_k \Theta^{i,M}_{T^2/Z_2,\eta}(z_k)\Theta^{j,M}_{T^2/Z_2,\eta}(z_k)\right) {\overline{(\chi^i_R)^{C}} \chi^j_R} + \text{h.c.},
\end{eqnarray}
%%%%%
{where} we can analytically obtain the effective Majorana mass matrix as
\begin{eqnarray}
\label{majoranamass}
(M_R)_{ij}=\sum_{k=1}^4h_k \Theta^{i,M}_{T^2/Z_2,\eta}(z_k)\Theta^{j,M}_{T^2/Z_2,\eta}(z_k){.}
\label{Maj}
\end{eqnarray}
{The} implicit factor $h_k (\mathcal{N}_M)^2$, which has mass dimension $+1$, provides a typical scale of Majorana masses. When $h_k / \sqrt{\mathcal{A}} \sim \mathcal{O}(1)$, this scale is close to the compactification scale $M_C$.

{Before {the end of} this section, we comment on the structures of the effective Majorana mass matrix.}
For several cases, we reach the formula for wave functions on the $Z_2$ fixed points,
\begin{eqnarray}
\label{fixedpointformula}
\Theta^{j,M}(-z_k)=(-1)^{M\delta_{k,4}}\Theta^{j,M}(z_k).
\end{eqnarray}
Following this formula and Eq.\,(\ref{Z2wavefunc}), we {find}
\begin{eqnarray}
\Theta^{j,M}_{T^2/Z_2,+1}(z_k)&=&\sqrt{2}\Theta^{j,M}(z_k),\\
\Theta^{j,M}_{T^2/Z_2,-1}(z_k)&=&0,
\end{eqnarray}
where $M$ is an even number or $k\neq4$.
In the case that $M$ is an odd number and $k=4$, we obtain
\begin{eqnarray}
\Theta^{j,M}_{T^2/Z_2,+1}(z_k)&=&0,\\
\label{fixedpointformula2}
\Theta^{j,M}_{T^2/Z_2,-1}(z_k)&=&\sqrt{2}\Theta^{j,M}(z_k).
\end{eqnarray}
%From Eq.\,(\ref{majoranamass}), a rank of this Majorana mass matrix is equal to the number of $Z_2$ fixed points where mass terms are {non-trivially} introduced.
%In other words, the number of non-zero $h_k\ (k=1,2,3,4)$ {naively} corresponds to the rank of the Majorana mass matrix.
%Hence the highest rank of it is four at most.
{The above properties are closely related to the rank of the Majorana mass matrix.}
%n addition to that,
{Eqs.}\,(\ref{fixedpointformula})\,--\,(\ref{fixedpointformula2}) imply that if $\eta=-1$ almost all values of wave functions on the $Z_2$ fixed points are vanishing and the highest rank of this case is one.
%In order to consider three-generation fermions, a rank of Eq.\,(\ref{majoranamass}) requires at least three,
{Thereby, we will {focus on} the case $\eta=+1$.}
%Note that a mass {scale} of this mass matrix is naively equal to the compactification scale of the extra dimensions.

%%%%%%%%%%%%%%%%%%%%%%%%%%%%%%%%%%%%%%%%%%%%%%%%%%%%%%
%%%%%%%%%%%%%%%%%%%%%%%%%%%%%%%%%%%%%%%%%%%%%%%%%%%%%%
\section{Seesaw scenario in magnetic compactifications}
%%%%%%%%%%%%%%%%%%%%%%%%%%%%%%%%%%%%%%%%%%%%%%%%%%%%%%
%%%%%%%%%%%%%%%%%%%%%%%%%%%%%%%%%%%%%%%%%%%%%%%%%%%%%%

%%%%%%%%%%%%%%%%%%%%%%%%%%%%%%%%%%%%
\subsection{The models}
%%%%%%%%%%%%%%%%%%%%%%%%%%%%%%%%%%%%

In this section, we consider the type I seesaw mechanism {under the} brane-localized {forms for} right-handed neutrino Majorana mass terms.
In our setup, the Dirac mass matrix drives from bulk Yukawa couplings among the leptons and the Higgs doublet.
This is a definitive difference from the model buildings in \cite{Buchmuller:2017vho,Buchmuller:2017vut} where the Yukawa couplings are also introduced to the fixed points.
A six-dimensional Lagrangian of our {scenario is summarized} as
\begin{eqnarray}
\mathcal{L}_{N}=-g\bar{L}NH - {\frac{1}{2}} \sum_{i=1}^4 h_i \, {\overline{(N_R)^C} N_R} \, \delta^2(z-z_i) + \text{h.c.},
\end{eqnarray}
where $L, N$ and $H$ are a six-dimensional left-handed lepton doublet, right-handed neutrino {singlet} and Higgs doublet, respectively.
{Same-sign 6D chiralities are} arranged for $L$ and $N$ to realize zero-mode left-handed neutrinos ($\nu_L$) from $L$ and right-handed ($\nu_R$) ones from $N$ with three generations (see Tab.\,\ref{fluxpattern}).\footnote{
{Possible 6D anomalies can be compensated by introducing additional 6D chiral matters without zero mode (see e.g.,~\cite{Dobrescu:2001ae}).
Another possibility {would be} to embed our phenomenological setup to a ten-dimensional super Yang--Mills theory.}
}
%They are satisfied the following chirality conditions,
%\begin{eqnarray}
%\Gamma^7L=-L, \qquad \Gamma^7N=+N,
%\end{eqnarray}
%where $\Gamma^7$ is a six-dimensional chirality operator,
%{where zero-mode left-handed ($\nu_L$) and right-handed ($\nu_R$) neutrinos with three generations are generated.}

In general, it is possible to consider multiple Higgs fields.
However, it is plausible that {the models with multiple Higgs doublets} are quite {uneasy} because they likely suffer from flavor changing neutral current(s), as well as the models have {many} parameters like the Higgs VEVs unless we concretely analyze the multiple Higgs potential.
Therefore, we focus on the case that the number of parameters is minimum, i.e., the generation of Higgs field is one.

In addition, we consider the  three generation of left- and right-handed neutrinos, where the definite number of the right-handed neutrinos has not been fixed yet.
According to the previous section, brane-localized fermions must be $Z_2$ even {($\eta=+1$)} and gauge invariance of Yukawa couplings demands $|M_1| + |M_2|=|M_3|$ for Case I and $|M_1| + |M_3| = |M_2|$ for the other cases, where $M_1, M_2,{\rm and}\ M_3$ denote the flux numbers for the Higgs doublet $H$, the left-handed lepton doublet $L$ and the right-handed neutrino $N$, respectively.
It is necessary to note that we need to interchange $M_2 \leftrightarrow M_3$ in using Yukawa couplings \eqref{yukawas} except for Case I.
This is because $|M_3|$ is {assumed to be the} maximal flux in the notation of \eqref{yukawas}.
Thus, flux configurations satisfying these conditions appear just in five patterns, as shown in Tab.\,\ref{fluxpattern}, where $\eta_1,\eta_2,{\rm and}\ \eta_3$ denote the $Z_2$ parities for {the Higgs doublet, the left-handed lepton doublet, and the right-handed neutrino}, respectively.

\begin{table}[t]
\begin{center}
\begin{tabular}{|c|ccc|ccc|}\hline
 & $\eta_1$ & $\eta_2$ & $\eta_3$ & $M_1$ & $M_2$ & $M_3$\\ \hline
Case I & $+1$ & $+1$ & $+1$ & $-1$ & $-4$ & $+5$ \\
Case II & $+1$ & $+1$ & $+1$ & $-1$ & $+5$ & $-4$ \\
Case III & $-1$ & $-1$ & $+1$ & $-3$ & $+7$ & $-4$ \\
Case IV & $-1$ & $-1$ & $+1$ & $-4$ & $+8$ & $-4$ \\
Case V & $-1$ & $-1$ & $+1$ & $-3$ & $+8$ & $-5$ \\\hline
\end{tabular}
\caption{{The five patterns of allowed model setups. Higgs field $H$ carries a flux number and $Z_2$ parity $(M_1, \eta_1)$. Similarly, left- and right-handed lepton $L$ and $N$ carry $(M_2, \eta_2)$ and $(M_3, \eta_3)$, respectively.}}
\label{fluxpattern}
\end{center}
\end{table}

{It should be noted about the Higgs VEV which causes the electroweak symmetry breaking.
To be naive, Eq.\,\eqref{KKscalar} implies that there is no massless mode in the scalar spectrum.
{A possible way to realize the EW scale would be to tune parameters in the Higgs potential.}
Another route for deriving a massless scalar in magnetized setups is to consider embeddings into more higher dimensional setup {with a larger gauge group}, for example, ten-dimensional super Yang--Mills theory, {where the scalar originates from a KK-decomposed higher-dimensional gauge boson} \cite{Abe:2012fj}.\footnote{
{See~\cite{Lim:2018lgg} for related discussions.
Quantum corrections in such setups are discussed in~\cite{Buchmuller:2016gib,Ghilencea:2017jmh}.}
}
Throughout our this paper, we assume that the Higgs massless mode causes the EW breaking {appropriately}.\footnote{When the 6D setup in this paper {can be} derived from some classes of superstring theory, in fact there are {promising} mechanisms that cause the EW breaking around $10^2$ GeV \cite{Ibanez:2006da, Abe:2015uma, Kobayashi:2015siy}.}}

After the Higgs boson develops its VEV $v = 174 \,\, {\rm GeV}$, we analytically express the Dirac neutrino mass matrix,
\begin{eqnarray}
(m_D)_{ij}=Y_{T^2/Z_2}^{ij} v.
\end{eqnarray}
Using this Dirac mass matrix and the right-handed Majorana mass matrix {in} Eq.\,(\ref{majoranamass}), the total neutrino mass matrix in the seesaw scenario is written as
\begin{eqnarray}
{\begin{pmatrix}
\overline{\nu_L} & \overline{\nu_R^c}
\end{pmatrix}
\begin{pmatrix}
0 &m_D\\[5pt]
m_D^\mathrm{T} &M_R
\end{pmatrix}
\begin{pmatrix}
\nu_L^c \\ \nu_R
\end{pmatrix}},
\end{eqnarray}
{where the indices for representing the three generations are suppressed.}
After all, we consider $M_R\gg m_D$ in an {ordinary} manner of the seesaw, and then the effective left-handed neutrino Majorana mass matrix can be described as
\begin{eqnarray}
\label{MLmatrix}
m_{LL} \simeq -m_DM_R^{-1}m_D^\mathrm{T}.
\end{eqnarray}

{Here, we should mention that additional contributions may occur through the seesaw mechanism as exchanges of KK neutrinos in the Majorana mass terms if the seesaw scale $M_R$ is close to the compactification scale $M_C$.
In this paper, we simply assume the relation $M_R \ll M_C$, which is realized by the condition $h_k / \sqrt{\mathcal{A}} \ll \mathcal{O}(1)$ to ignore such contributions, for simplicity (refer to the {sentences} around Eq.~(\ref{majoranamass})).}

\subsection{Numerical analyses}
In the following, we will analyze the relations between model parameters and several experimental data.
In our setup, there are apparently {seven} real model parameters, i.e., the complex structure modulus {$\tau \in \mathbb{C}$, the overall Yukawa coupling $g$,} and localized masses on fixed points $h_k \,\, (k=1,2,3,4)$.
%\footnote{The compactification scale ($\sim 1/\sqrt{\cal A}$) can be always {chosen to give an appropriate seesaw scale which is calculated as the seesaw scale ($\sim {\cal O}(h_k) M_C$).} However, in this paper, we ignore this degree of freedom by focusing on the ratio of neutrino mass squared differences $r\equiv \Delta m^2_{21}/\Delta^2_{31}$.}
For scanning such model parameters, we try to fit the three lepton mixing angles $\theta_{ij} \,\, (ij=12, 23, 13)$, the CP violating phase $\delta_{\rm CP}$, and the ratio of {two} mass squared differences of {the observed neutrino states} $r$.
%Note that the total neutrino mass scale like eV scale, can be always realized by fine-tuning the compactification scale.
%Thus, we will not discuss this topic in detail in this paper.
In our models, these experimental values are independent of an overall factor of the mass matrix (\ref{MLmatrix}).
Therefore, except for the configurations of magnetic fluxes {and $Z_2$ parities}, {effective degrees of freedom for describing mixing structures and mass differences} in our models are complex structure modulus $\tau\in\mathbb{C}$ and the ratio of brane-localized mass parameters $\rho_{{k'}} \equiv h_{{k'}}/h_1\in\mathbb{R} \,\, ({k'} = 2,3,4)${,} if we set $h_1 \neq 0$ {and $g \neq 0$} and {assume that a sub-eV typical neutrino scale is generated by a suitable relationship between $h_1$ and $g$ (through the type I seesaw mechanism).}
In the following analyses, we set $h_1 \neq 0$ and $g \neq 0$.

It is convenient to show a numerical sample of matrix patterns in the Dirac and Majorana mass matrices.
For $\tau = i$ and $\rho_2 = \rho_3 = 1$ in Case I, they are given as
\begin{gather}
m_D \propto 
\begin{pmatrix}
1.12 & 0.13 & 7 \times 10^{-5} \\
0.03 & 0.96 & 0.27 \\
3 \times 10^{-7} & 0.0056 & 0.85
\end{pmatrix}v,
\\
M_R \propto 
\begin{pmatrix}
6.33 & 0.043 & 0.88 \\
0.043 & 3.98 & 1.34 \\
0.887 & 1.34 & 4.70  
\end{pmatrix}M_C
.
\end{gather}
In this parameter {pattern}, it is found that diagonal elements are dominant and all elements are real.
Since the two of three neutrinos have a large mixing in the right-handed Majorana mass matrix, it can be promising in explaining the observed neutrino large mixings.
For another value of $\tau = 1 + i$, we obtain
\begin{gather}
m_D \propto 
\begin{pmatrix}
 1.12 & -0.104+0.076 i & - 5 \times 10^{-5} - 4 \times 10^{-5} i \\
 -0.022-0.022 i & 0.95 \, +0.15 i & 0.043 \, +0.27 i \\
 -3 \times 10^{-7} & 0.0045 \, -0.0033 i & 0.69 \, +0.50 i
\end{pmatrix}v,
\\
M_R \propto 
\begin{pmatrix}
 6.32\, +0.0049 i & 0.025 \, -0.035 i & -0.67 + 0.30 i \\
 0.025 \, -0.035 i & 0.76\, +3.54 i & 0.024 \, +1.32 i \\
 -0.67+0.30 i & 0.024\, +1.32 i & 4.42\, +1.35 i
\end{pmatrix}M_C.
\end{gather}
It is easy to find that absolute value of each entry is {almost} the same as that before.
However, complex phases appear in several elements.
Hence, non-zero value of ${\rm Re} \, \tau$ may fit a CP violating phase, as shown in the quark sector \cite{Kobayashi:2016qag}.

Now, we analyze the left-handed Majorana matrix shown in the previous subsection. From Eq.\,(\ref{MLmatrix}), we will derive the lepton mixing angles, the CP violating phase, and the ratio for mass squared differences of neutrino masses by numerical calculations.
Then, we will search parameter regions for reproducing neutrino experimental data \cite{Esteban:2016qun} as systematically as possible.\footnote{{Here, we assume that the charged lepton sector does not disturb patterns of neutrino mass matrix $m_{LL}$.
%Apparently this assumption looks strange, however, in fact
{It is quite reasonably justified in what follows.}
In the charged lepton sector as well as quark sectors, the mass matrix has relatively small off diagonal entries in contrast to diagonal entries {to reproduce the hierarchical mass differences} as shown in {e.g.,} {Subsection}\, 4.1 of \cite{Abe:2015yva}.
{This means} that contributions from the charged lepton mass matrix {may be} estimated to be typically small.
For the reason, we evaluate the {lepton} mixing angles only from the neutrino sector.}}
The lepton mixing matrix, called Pontecorvo--Maki--Nakagawa--Sakata (PMNS) matrix $U_{\rm PMNS}$, is conventionally written as
\begin{eqnarray}
U_{\rm PMNS}=
 \begin{pmatrix}
c_{12}c_{13} &s_{12}c_{13} &s_{13}e^{-i\delta_{\rm CP}}\\
-s_{12}c_{23}-c_{12}s_{23}s_{13}e^{i\delta_{\rm CP}} &c_{12}c_{23}-s_{12}s_{23}s_{13}e^{i\delta_{\rm CP}} &s_{23}c_{13}\\
s_{12}s_{23}-c_{12}c_{23}s_{13}e^{i\delta_{\rm CP}} &-c_{12}s_{23}-s_{12}c_{23}s_{13}e^{i\delta_{\rm CP}} &c_{23}c_{13} 
 \end{pmatrix}
\begin{pmatrix}
e^{i\alpha} &0 &0\\
0 &e^{i\beta} &0\\
0 &0 &1
\end{pmatrix}
,\nonumber\\
\end{eqnarray}
where $s_{ij}$ and $c_{ij}$ {($s_{ij},\, c_{ij} > 0$)} denote $\sin\theta_{ij}$ and $\cos\theta_{ij}$, $\delta_{\rm CP}$ denotes the CP violating phase and $\alpha$ and $\beta$ are the Majorana phases.
In our numerical calculations, {we target} the $3\sigma$-favored ranges of the lepton mixing angles and mass squared differences of neutrino masses {which were derived through the global fit in} \cite{Esteban:2016qun},
\begin{eqnarray}
\label{evalangle}
0.271<\sin^2\theta_{12}<0.345, \quad 0.385<\sin^2\theta_{23}<0.635, \quad 0.01934<\sin^2\theta_{13}<0.02392,
\end{eqnarray}
\begin{eqnarray}
\label{evalmass}
7.03<\frac{\Delta m_{21}^2}{10^{-5} \, {\rm eV}^2}<8.09, \qquad 2.407<\frac{\Delta m_{3 \ell}^2 \, {(= \Delta m_{31}^2)}}{10^{-3} \, {\rm eV}^2}<2.643,
\end{eqnarray}
with $\Delta m^2_{ij} = m_i^2 - m_j^2$ and the normal mass hierarchy {(NH)} being assumed.
The mass ratio $r$ is defined as $\Delta m^2_{21} / |\Delta m^2_{3\ell}|$.
{The mass difference $\Delta m^2_{3\ell}$ is defined as $\Delta m^2_{3\ell} \equiv \Delta m_{31}^2 \, (>0)$ for NH and $\Delta m^2_{3\ell} \equiv \Delta m_{32}^2 \, (<0)$ for the inverted hierarchy~(IH)~\cite{Esteban:2016qun}.}
In addition to these observables, a promising $1\sigma$ range of the CP violating phase has been recently measured by many neutrino experiments and an experimental values \cite{Esteban:2016qun} for NH is known as~\footnote{
{The $3\sigma$ favored ranges of $\delta_{\rm CP}$ and $r$ in the NH case reported in Ref.~\cite{Esteban:2016qun} are as follows:
$0^\circ \leq \delta_{\rm CP} \leq 360^\circ$,
$-1.57 \leq \log_{10}{r} \leq -1.49$,
where the latter is evaluated from Eq.~(\ref{evalmass}).}
}
\begin{eqnarray}
\label{evalcp}
202^\circ<\delta_{\rm CP}<312^\circ.
\end{eqnarray}
{We note that similar analyses made by different groups have been {also reported recently}~\cite{Capozzi:2017ipn,Gariazzo:2018uwn}.}

\begin{figure}[t]
\centering
\includegraphics[clip, width=0.43\columnwidth]{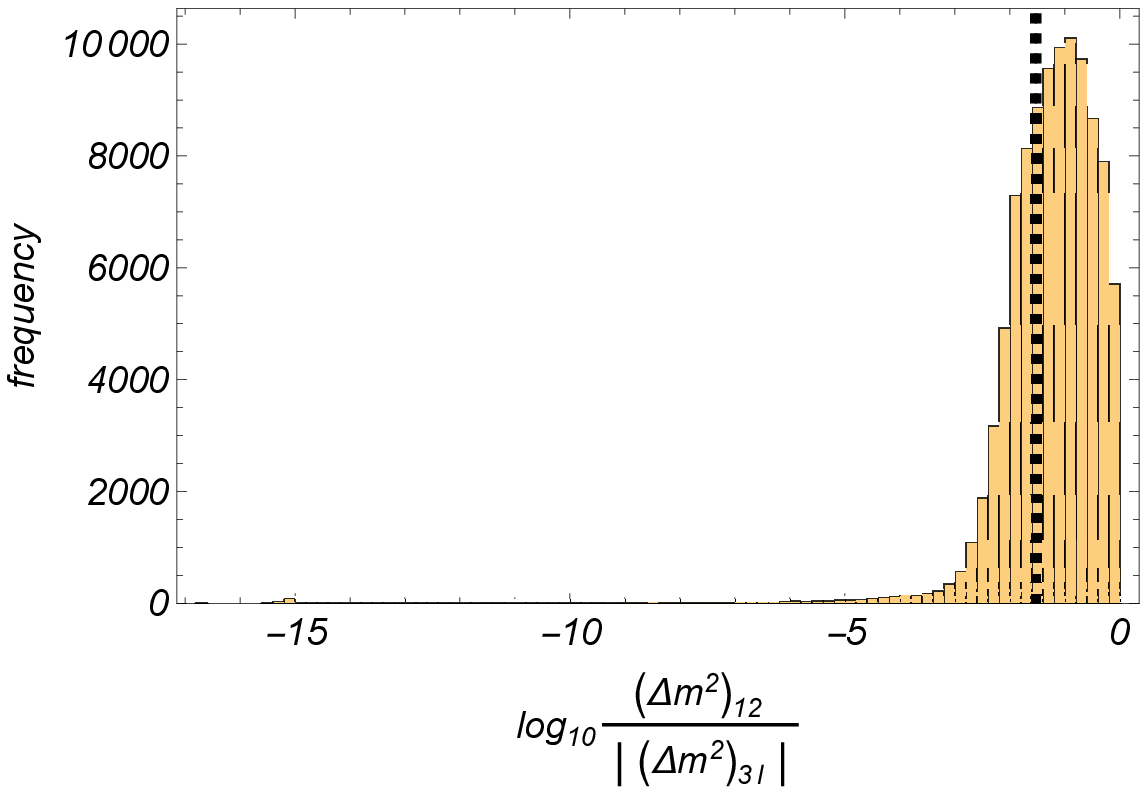} \quad 
\includegraphics[clip, width=0.43\columnwidth]{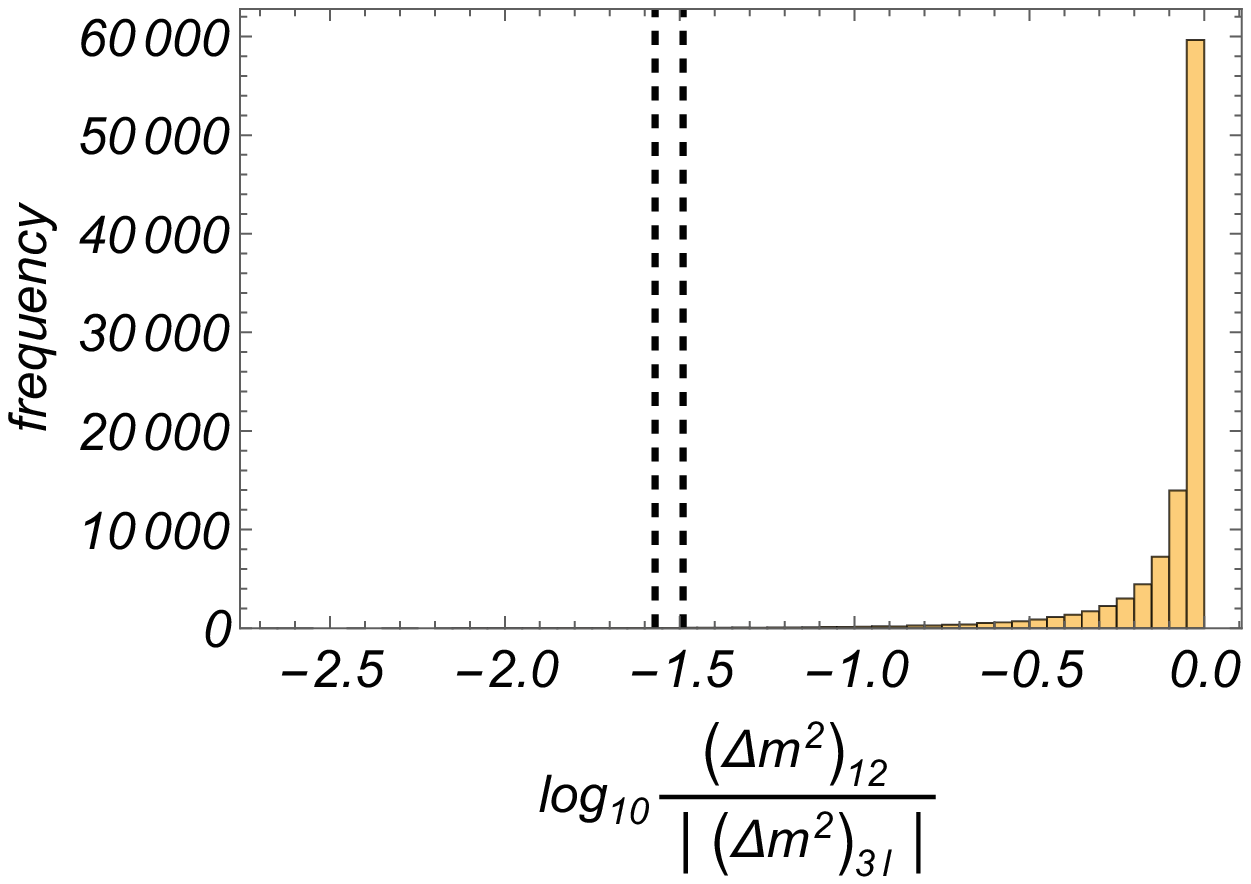}
\caption{Distributions of the ratio $\log_{10}(\Delta m^2_{21}/|\Delta m^2_{3\ell}| )$ under the assumptions of NH (left panel) and IH (right panel) in Case I. Here, we impose no cut for the three mixing angles within the $3\sigma$ ranges.
{In the panels, the regions between the two vertical dashed black lines are $3\sigma$-favored for $\log_{10}(\Delta m^2_{21}/|\Delta m^2_{3\ell}| )$ \cite{Esteban:2016qun}.}
{Here we take $10^5$ points for each plot.}
}
\label{fig:NH_vs_IH}
\end{figure}

In this paper, we will analyze only the NH case since our model cannot reproduce all of observed data, especially mass squared differences in the {IH} case {(see Fig.~\ref{fig:NH_vs_IH})}.
As discussed in \cite{Abe:2008sx}, there are just five patterns for the configurations of magnetic fluxes which can generate appropriate Dirac mass matrices with three-generation leptons (see Tab.\,\ref{fluxpattern}).
Thus, we will numerically search parameter regions to reproduce the experimental data (\ref{evalangle}) {and} (\ref{evalmass}){;} and {also} (\ref{evalcp}) {(if possible)} for all five patterns.
%However, Case III and Case V do not have any parameter regions for realizing experimental data, hence we discuss only the other three patterns.

In Case I {and Case V}, the flux for the right-handed neutrino is an odd integer, then a brane-mass parameter $h_4$ do not affect computations as we discussed in the previous section. 
In other words, even if the brane-mass parameter is non-zero, the Majorana mass is {not changed {in} the fourth fixed point ($k=4$)}.
Therefore, free parameters for this pattern are $\tau$, $\rho_2$ and $\rho_3$. 
On the other hand, in the other cases, the flux for the right-handed neutrino is an even integer, {and} then a brane-mass parameter $h_4$ affects computations.
Therefore, free parameters for these patterns are $\tau$, $\rho_2$, $\rho_3$ and $\rho_4$.
In these conditions, we set inputs of free parameters {as shown in} Tab.\,\ref{paramlist} and the results are shown in Tab.\,\ref{resultlist}. These results are in the {$3\sigma$-favored region} of all experimentally observed data~{\cite{Esteban:2016qun}}.
In the next subsection, we will show some details of numerical analyses.

\begin{table}[t]
\begin{center}
\begin{tabular}{|c|ccccc|c|}\hline
 & Case I & Case II & Case III & Case IV & Case V & Central value~\cite{Esteban:2016qun} \\ \hline
$\sin^2\theta_{12}$ & $0.306$ & $0.291$ & $0.345$ & $0.284$ & $0.337$ & $0.306$\\
$\sin^2\theta_{23}$ & $0.512$ & $0.520$ & $0.490$ & $0.453$ & $0.480$ & $0.441$\\
$\sin^2\theta_{13}$ & $0.0194$ & $0.0237$ & $0.0238$ & $0.0224$ & $0.0211$ & $0.02166$\\
$\delta_{\rm CP}$ & $84.8^\circ$ & $350^\circ$ & $325^\circ$ & $27.2^\circ$ & $289^\circ$ & $261^\circ$\\
$\log_{10}\left( \frac{\Delta m^2_{21}}{|\Delta m^2_{3\ell}|} \right) = \log_{10}{r}$ 
                             & $-1.52$ & $-1.53$ & $-1.52$ & $-1.56$ & $-1.51$ & $-1.53$\\ \hline
\end{tabular}
\caption{The results for the mixing angles, CP phases, and mass squared differences.}
\label{resultlist}
\end{center}
\end{table}

\begin{table}[t]
\begin{center}
\begin{tabular}{|c|ccccc|}\hline
 & Case I & Case II & Case III & Case IV & Case V \\ \hline
Re\,$\tau$ & $1.748$ & $-2.246$ & $1.114$ & $0.2800$ & $2.652$ \\
Im\,$\tau$ & $0.04900$ & $1.432$ & $0.9880$ & $1.059$ & $0.8210$ \\
$\rho_2$ & $-0.59$ & $0.38$ & $0.34$ & $0.32$ & $0.90$ \\
$\rho_3$ & $0.21$ & $0.35$ & $-0.40$ & $0.60$ & $-0.69$ \\
$\rho_4$ & {$0$} & $-0.23$ & $0.88$ & $-0.28$ & {$0$} \\ \hline
\end{tabular}
\caption{Input parameters ($\tau,\rho_2,\rho_3,\rho_4$) for generating the configurations in Table~\ref{resultlist}.
{Note that $\rho_4$ is ineffective in Case I and Case V and then we set zero for $\rho_4$ in these cases.}}
\label{paramlist}
\end{center}
\end{table}

\begin{figure}[t]
\centering
\includegraphics[clip, width=0.40\columnwidth]{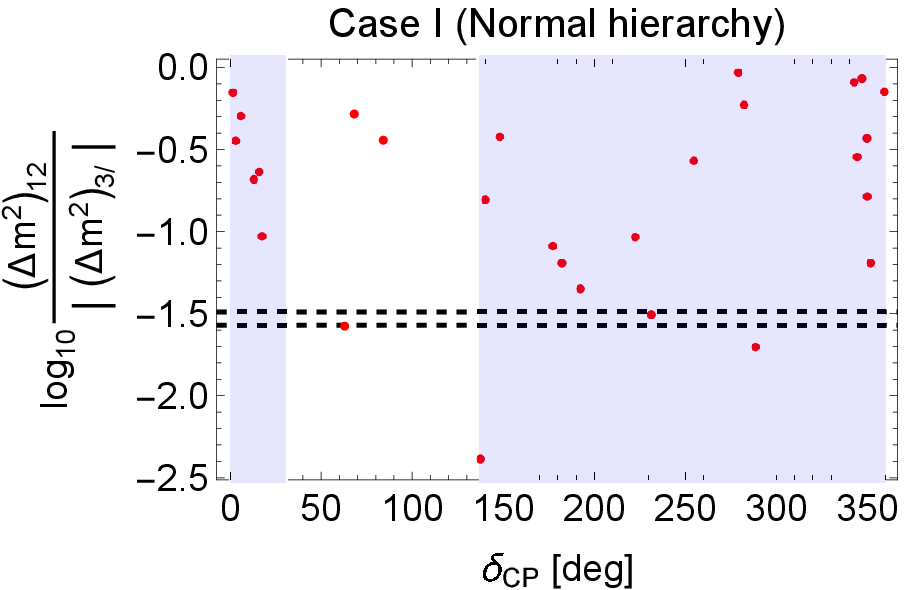} \qquad
\includegraphics[clip, width=0.40\columnwidth]{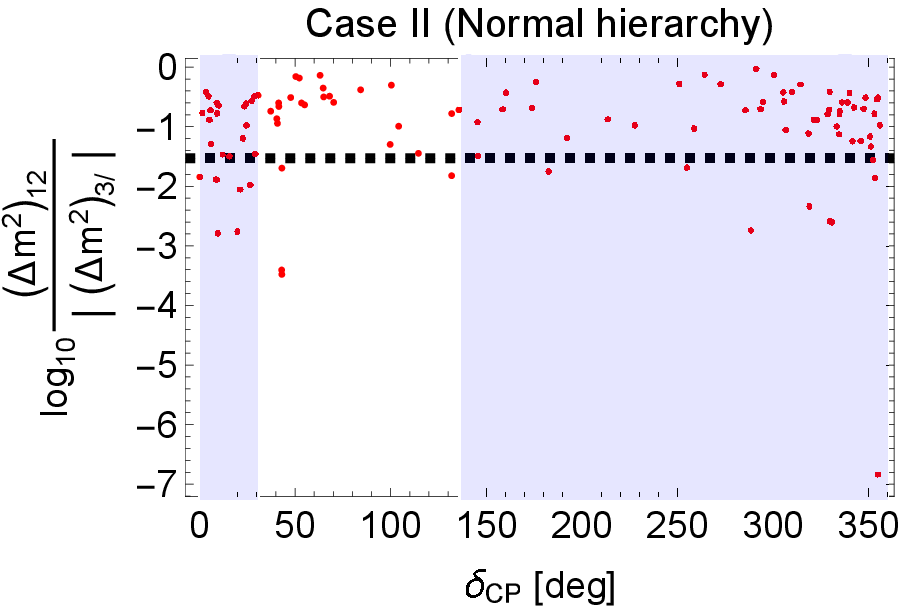} \\[15pt]
\includegraphics[clip, width=0.40\columnwidth]{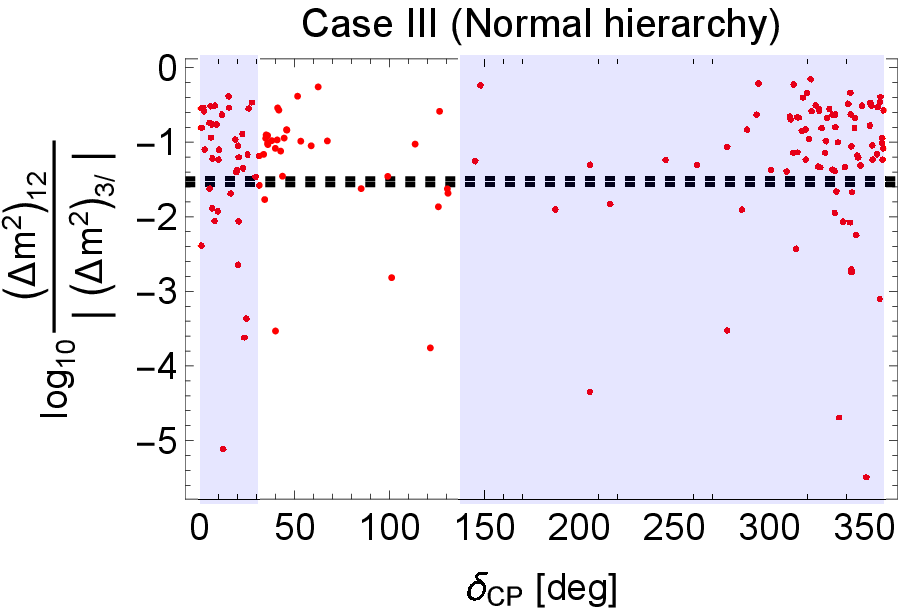} \qquad
%%%
%\includegraphics[clip, width=0.40\columnwidth]{plot_Case_4.eps} \\[15pt]
%\includegraphics[clip, width=0.40\columnwidth]{plot_Case_5.eps}
\includegraphics[clip, width=0.40\columnwidth]{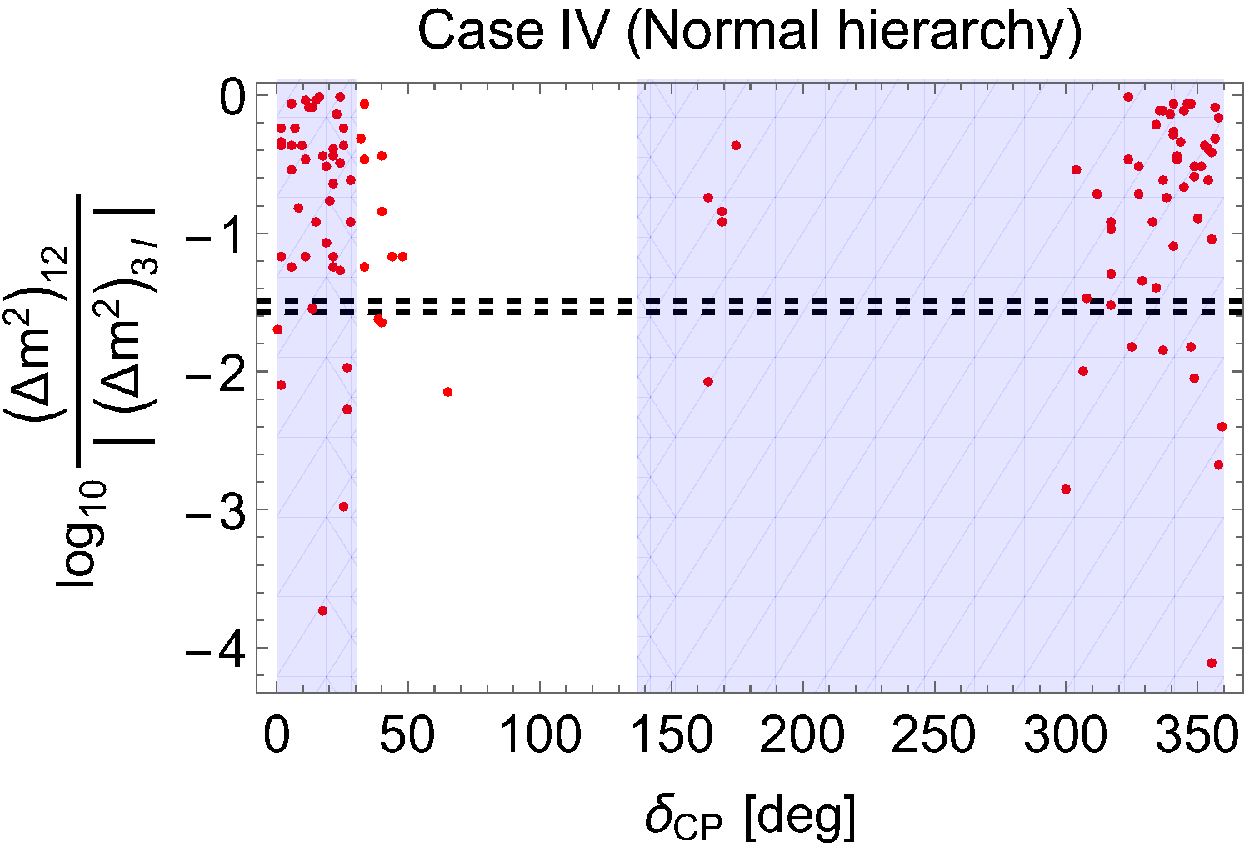} \\[15pt]
\includegraphics[clip, width=0.40\columnwidth]{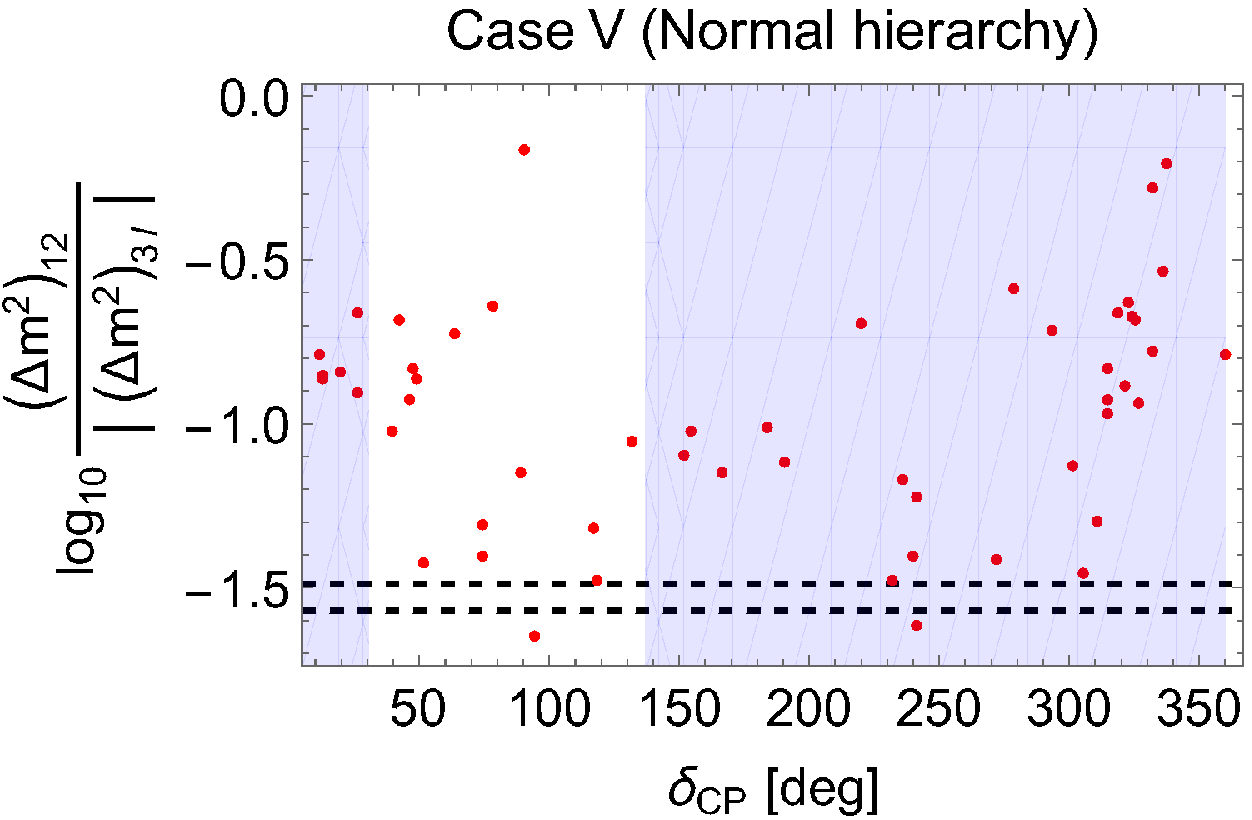} 
\caption{{$\log_{10}(\Delta m^2_{21}/|\Delta m^2_{3\ell}| )$\,--\,$\delta_\text{CP}$ distributions among the parameter points where the three mixing angles are within the $3\sigma$ regions.
The pale blue regions show the $3\sigma$-favored region of $\delta_\text{CP}$ in a recent global fit~\cite{Capozzi:2017ipn}.
It is noted that no region of $\delta_\text{CP}$ is $3\sigma$-disfavored in the results of the recent global fits \cite{Esteban:2016qun,Gariazzo:2018uwn}.
In each plot, the region between the two horizontal dashed black lines are $3\sigma$-favored for $\log_{10}(\Delta m^2_{21}/|\Delta m^2_{3\ell}| )$ \cite{Esteban:2016qun}.}}
\label{fig:3sigma_dist}
\end{figure}

\subsection{{Detailed analyses of each case}}
%Next, we consider the stability of the input parameters numerically. For example, in the case II, Re$\tau$, Im$\tau$, $\rho_2$, $\rho_3$, and $\rho_4$ are chosen randomly in the values in (Tab.~\ref{paramlist})$\pm0.1$ with $10^4$ trials and we draw histograms. The orange histograms show results in the $3\sigma$ range. In these result, about one-tenth of these calculations reproduces the experimental data (\ref{evalangle}), (\ref{evalmass}), and (\ref{evalcp}).
%{\color{red} This part has not been modified. We are waiting for Nishiwaki-san's crosscheck.}

{At first, we look whether the NH or IH case is preferred in our seesaw texture originating from the magnetized extra dimension with orbifolding.
In Fig.\,\ref{fig:NH_vs_IH}, we show the distributions of the ratio defined as $\log_{10}(\Delta m^2_{21}/|\Delta m^2_{3\ell}| )$ in Case I, where NH and IH are assumed in the left and right panels, respectively.
Here, we impose no cut for the three mixing angles within the $3\sigma$ ranges~{\cite{Esteban:2016qun}}.
We immediately recognize that IH is highly disfavored since the corresponding range calculated from the global fit result in \cite{Esteban:2016qun} is located far away from the peak of the {obtained} distribution.
{We found that the other cases have similar properties to Case I, {where we can conclude} that the IH case is disfavored in any case.}
Thereby hereafter, we only focus on the NH case in the five cases.

Next, we impose the $3\sigma$ conditions on the three mixing angles on the randomly generated configurations from the scattered parameters within the designated ranges as
\begin{align}
\text{Re} \, \tau \in [-\pi, \pi], \qquad
\text{Im} \, \tau \in (0, \pi], \qquad
\rho_{2,3,4} \in [-1,-0.1] \cup [0.1, 1],
\end{align}
where $10^6$ points are taken into account in each case individually.
The correlations between $\log_{10}(\Delta m^2_{21}/|\Delta m^2_{3\ell}| )$ and $\delta_\text{CP}$ are described in Fig\,\ref{fig:3sigma_dist}, where a few points {(in each case)} are $3\sigma$ acceptable also in the two values even though we take the severest result of such global fits among the one reported recently~\cite{Capozzi:2017ipn}.
Also, we explicitly provide {a sample} in every case, summarized in Tab.~\ref{resultlist}, where they are generated when we adopt the parameters shown in Tab.~\ref{paramlist}.
It is noted that the first four/two digits of $\tau$/$\rho_{2,3,4}$ looks sensitive to results in general.}

In the rest of this section, we {make a comment} on a {possible} correlation between ${\rm Re} \, \tau$ and $\delta_{\rm CP}$.
{As shown in} Fig.\,3, we obtained only the {$\sim 15$} number of candidates for allowed parameter {points in total}.
This is due to the fact that the mass matrices {stemming} from \eqref{yukawas} and \eqref{Maj} contain 
the Jacobi theta function that is defined in terms of an infinite summation over {integers}.
Evaluating values of the function takes considerable time.
{Also, values of the Jacobi theta function are very sensitive to $\text{Im}\,\tau$ roughly like $e^{- c \, \text{Im}\,\tau}$ with a constant $c$. Therefore, we should carefully investigate effects originating from slight differences in $\text{Im}\,\tau$.}
For {these difficulties in calculation time}, we {might ought to} conclude that it is {very difficult} to extract strong predictions in {the distributions of the realized CP phase concretely with keeping the current accuracy in the realized values of experimental measurements}.
However, we can {get a clue for} qualitative understanding for the CP phase {through the following speculation}.
In the light of a previous study \cite{Kobayashi:2016qag}, one {finds} that {the} real part of 
{the} complex structure modulus {generates non-zero physical} values of the CP phase.
{In mass matrices} stemming from flux compactification in quarks,  {we also observed considerable correlations between the value of ${\rm Re} \, \tau$ and the resulting (quark) CP violation phase}.
From {these observations}, we {can make a suggestion} that an observed region of neutrino CP phase at {a} confidence level {may} restrict an allowed range of ${\rm Re} \, \tau$.
{This point would lead to putting a constraint on types of possible mechanisms for moduli stabilization.}
It would be possible to reach strong predictions for the CP phase by combining our setups and appropriate moduli stabilizations in a future project.
Nevertheless, being different from the quark case without brane-local term discussed in~\cite{Kobayashi:2016qag}, the existence of the brane-local Majorana mass terms may lead to more complex pattens in the distributions of the realized CP angle, even though the coefficients of the mass terms are real as we assumed.
Thereby, an exhaustive calculation with a very considerable calculation cost would be required for unveiling possible hidden patterns of the CP angle in the current system and we do not explore the detail of it in this manuscript.

\section{Conclusion}

{In this paper, we have explored a new avenue to a natural explanation of the observed tiny neutrino masses with a dynamical realization of the three-generation structure in the neutrino sector.
Under the magnetized background, matters have multiply-degenerated zero modes and the whole intergenerational structures (before diagonalization of mass matrices) are dynamically determined.
In this sense, we can conclude that our scenario is favored in the concept of minimality, where no degree of freedom remains to deform part of an intergenerational structure by hand freely.

Another good feature in our story is that only one Higgs doublet is enough for reproducing measured configurations of neutrinos, being different from the case of the quarks which have been discussed in various previous works.
On magnetized $T^2/Z_2$ orbifolds, four fixed points are observed, where we can write down Majorana-type mass terms of {an} $SU(2)_L$ singlet neutrino field, with different coefficients.
Our numerical calculations have clarified that to find acceptable parameter configurations (where the three mixing angles and the mass ratio are within the $3\sigma$ ranges) is not exceedingly tough.
As shown in Fig.~\ref{fig:3sigma_dist}, after a $10^6$-time random scan, a few valid cases are excavated in all of the five reasonable configurations in the magnetic fluxes and $Z_2$ parities (as summarized in Tab.~\ref{fluxpattern}).

Due to the complexity of the theta function and $Z_2$ orbifolding, physical CP-violating phase is realized~\cite{Lim:1990bp,Lim:2009pj,Kobayashi:2016qag}.
When the Dirac CP phase in the Pontecorvo--Maki--Nakagawa--Sakata matrix is measured much more precisely,
we may clarify what type of the fluxes and $Z_2$ parities is more favorable.
Allowing complex coefficients in the brane-localized Majorana-type mass terms may lead to a successful leptogenesis scenario~\cite{Fukugita:1986hr}, where details of such a possibility can be discussed in a separated publication in future.}

{Before closing this section, we comment on {the} $U(1)$ gauge symmetry and its breaking that we have used to obtain the family structure in leptons.
Since we focus only on the neutrino sector in this paper, we cannot decide whether such $U(1)$ symmetry is anomaly free or not in principle.
The fate of {the} $U(1)$ symmetry would {highly depend on philosophies in} model embeddings.
For example, {multiple} $U(1)$ symmetries are used in \cite{Abe:2012fj}, and it is well known that multiple $U(1)$ symmetries appear even in the intersecting D-brane scenario \cite{Ibanez:2012zz} (and references therein).
Even if the $U(1)$ is anomalous, there are possibilities to cancel it via the Green--Schwartz mechanism \cite{Green:1984sg} in the case that {the present scenario is realized by a} more fundamental theory, e.g., the superstring theory.
%In the other cases, the $U(1)$ may be broken by some effects in a framework of the 4D effective theory.
{Another clue for breaking the $U(1)$ gauge symmetry is to add a scalar field for a spontaneous breakdown.}
In other words, the $U(1)$ breaking can be concluded after the total {ultraviolet-completed} setups are identified.
Although we do not identify the total setups, the results we obtained {can be} typical patterns of neutrino mixings in flux compactification of toroidal orbifolds.}

%\bigskip
%In this paper, we have considered a six-dimensional theory that the geometry of extra dimensions is a toroidal orbifold $T^2/Z_2$ and non-zero magnetic fluxes.
%On such a geometry, we have added brane-localized terms at $Z_2$ fixed points.
%In this theory, we have calculated four-dimensional effective theory by the dimensional reduction, then gotten the right-handed Majorana mass matrix from the brane-localized terms and the neutrino Dirac mass matrix from the overlap integral of wave functions of the extra dimension, and then analyzed four-dimensional neutrino mass matrices. In this case, there have been  just five patterns of fluxes which induce three generation matter fermions, one generation Higgs and an right-handed Majorana mass matrix that a rank is more than or equal to three. In this scenario, we have obtained the Dirac and Majorana mass matrices analytically. Then, we have searched parameter regions that reproduce the experimental data for all five patterns of fluxes, especially for the NH case. In the result, in the three of five cases, we have obtained well parameter regions that can induce the experimental data.

\section*{Acknowledgement}
The authors would like to thank Hiroyuki Abe for valuable comments.
Y.T. would like to thank Wilfried Buchm\"uller for comments and suggestions on this manuscript.
Y.T. is supported in part by Grants-in-Aid for Scientific Research No.~16J04612 from the Ministry of Education, Culture, Sports, Science and Technology in Japan.

%0
%\bibliographystyle{junsrt}
\bibliographystyle{utphys}
\bibliography{references}

\providecommand{\href}[2]{#2}\begingroup\raggedright\begin{thebibliography}{10}

\bibitem{Aad:2012tfa}
{\bfseries ATLAS} Collaboration, G.~Aad {\em et~al.}, ``{Observation of a new
  particle in the search for the Standard Model Higgs boson with the ATLAS
  detector at the LHC},''
  \href{http://dx.doi.org/10.1016/j.physletb.2012.08.020}{{\em Phys.Lett.}
  {\bfseries B716} (2012) 1--29},
\href{http://arxiv.org/abs/1207.7214}{{\ttfamily arXiv:1207.7214 [hep-ex]}}.
%%CITATION = ARXIV:1207.7214;%%.

\bibitem{Chatrchyan:2012ufa}
{\bfseries CMS} Collaboration, S.~Chatrchyan {\em et~al.}, ``{Observation of a
  new boson at a mass of 125 GeV with the CMS experiment at the LHC},''
  \href{http://dx.doi.org/10.1016/j.physletb.2012.08.021}{{\em Phys.Lett.}
  {\bfseries B716} (2012) 30--61},
\href{http://arxiv.org/abs/1207.7235}{{\ttfamily arXiv:1207.7235 [hep-ex]}}.
%%CITATION = ARXIV:1207.7235;%%.

\bibitem{Abe:2013hdq}
{\bfseries T2K} Collaboration, K.~Abe {\em et~al.}, ``{Observation of Electron
  Neutrino Appearance in a Muon Neutrino Beam},''
  \href{http://dx.doi.org/10.1103/PhysRevLett.112.061802}{{\em Phys. Rev.
  Lett.} {\bfseries 112} (2014) 061802},
\href{http://arxiv.org/abs/1311.4750}{{\ttfamily arXiv:1311.4750 [hep-ex]}}.
%%CITATION = ARXIV:1311.4750;%%.

\bibitem{Adamson:2016tbq}
{\bfseries NOvA} Collaboration, P.~Adamson {\em et~al.}, ``{First measurement
  of electron neutrino appearance in NOvA},''
  \href{http://dx.doi.org/10.1103/PhysRevLett.116.151806}{{\em Phys. Rev.
  Lett.} {\bfseries 116} no.~15, (2016) 151806},
\href{http://arxiv.org/abs/1601.05022}{{\ttfamily arXiv:1601.05022 [hep-ex]}}.
%%CITATION = ARXIV:1601.05022;%%.

\bibitem{Patrignani:2016xqp}
{\bfseries Particle Data Group} Collaboration, C.~Patrignani {\em et~al.},
  ``{Review of Particle Physics},''
\href{http://dx.doi.org/10.1088/1674-1137/40/10/100001}{{\em Chin. Phys.}
  {\bfseries C40} no.~10, (2016) 100001}.
%%CITATION = CHPHD,C40,100001;%%.

\bibitem{Minkowski:1977sc}
P.~Minkowski, ``{$\mu \to e\gamma$ at a Rate of One Out of $10^{9}$ Muon
  Decays?},''
\href{http://dx.doi.org/10.1016/0370-2693(77)90435-X}{{\em Phys. Lett.}
  {\bfseries 67B} (1977) 421--428}.
%%CITATION = PHLTA,67B,421;%%.

\bibitem{Yanagida:1979as}
T.~Yanagida, ``{HORIZONTAL SYMMETRY AND MASSES OF NEUTRINOS},''
{\em Conf. Proc.} {\bfseries C7902131} (1979) 95--99.
%%CITATION = CONFP,C7902131,95;%%.

\bibitem{GellMann:1980vs}
M.~Gell-Mann, P.~Ramond, and R.~Slansky, ``{Complex Spinors and Unified
  Theories},'' {\em Conf. Proc.} {\bfseries C790927} (1979) 315--321,
\href{http://arxiv.org/abs/1306.4669}{{\ttfamily arXiv:1306.4669 [hep-th]}}.
%%CITATION = ARXIV:1306.4669;%%.

\bibitem{Mohapatra:1979ia}
R.~N. Mohapatra and G.~Senjanovic, ``{Neutrino Mass and Spontaneous Parity
  Violation},''
\href{http://dx.doi.org/10.1103/PhysRevLett.44.912}{{\em Phys. Rev. Lett.}
  {\bfseries 44} (1980) 912}.
%%CITATION = PRLTA,44,912;%%.

\bibitem{Schechter:1980gr}
J.~Schechter and J.~W.~F. Valle, ``{Neutrino Masses in SU(2) x U(1)
  Theories},''
\href{http://dx.doi.org/10.1103/PhysRevD.22.2227}{{\em Phys. Rev.} {\bfseries
  D22} (1980) 2227}.
%%CITATION = PHRVA,D22,2227;%%.

\bibitem{Froggatt:1978nt}
C.~D. Froggatt and H.~B. Nielsen, ``{Hierarchy of Quark Masses, Cabibbo Angles
  and CP Violation},''
\href{http://dx.doi.org/10.1016/0550-3213(79)90316-X}{{\em Nucl. Phys.}
  {\bfseries B147} (1979) 277--298}.
%%CITATION = NUPHA,B147,277;%%.

\bibitem{Ishimori:2010au}
H.~Ishimori, T.~Kobayashi, H.~Ohki, Y.~Shimizu, H.~Okada, and M.~Tanimoto,
  ``{Non-Abelian Discrete Symmetries in Particle Physics},''
  \href{http://dx.doi.org/10.1143/PTPS.183.1}{{\em Prog. Theor. Phys. Suppl.}
  {\bfseries 183} (2010) 1--163},
\href{http://arxiv.org/abs/1003.3552}{{\ttfamily arXiv:1003.3552 [hep-th]}}.
%%CITATION = ARXIV:1003.3552;%%.

\bibitem{ArkaniHamed:1999dc}
N.~Arkani-Hamed and M.~Schmaltz, ``{Hierarchies without symmetries from extra
  dimensions},'' \href{http://dx.doi.org/10.1103/PhysRevD.61.033005}{{\em Phys.
  Rev.} {\bfseries D61} (2000) 033005},
\href{http://arxiv.org/abs/hep-ph/9903417}{{\ttfamily arXiv:hep-ph/9903417
  [hep-ph]}}.
%%CITATION = HEP-PH/9903417;%%.

\bibitem{Bachas:1995ik}
C.~Bachas, ``{A Way to break supersymmetry},''
\href{http://arxiv.org/abs/hep-th/9503030}{{\ttfamily arXiv:hep-th/9503030
  [hep-th]}}.
%%CITATION = HEP-TH/9503030;%%.

\bibitem{Cremades:2004wa}
D.~Cremades, L.~Ibanez, and F.~Marchesano, ``{Computing Yukawa couplings from
  magnetized extra dimensions},''
  \href{http://dx.doi.org/10.1088/1126-6708/2004/05/079}{{\em JHEP} {\bfseries
  0405} (2004) 079},
\href{http://arxiv.org/abs/hep-th/0404229}{{\ttfamily arXiv:hep-th/0404229
  [hep-th]}}.
%%CITATION = HEP-TH/0404229;%%.

\bibitem{Braun:2006se}
A.~P. Braun, A.~Hebecker, and M.~Trapletti, ``{Flux Stabilization in 6
  Dimensions: D-terms and Loop Corrections},''
  \href{http://dx.doi.org/10.1088/1126-6708/2007/02/015}{{\em JHEP} {\bfseries
  02} (2007) 015},
\href{http://arxiv.org/abs/hep-th/0611102}{{\ttfamily arXiv:hep-th/0611102
  [hep-th]}}.
%%CITATION = HEP-TH/0611102;%%.

\bibitem{Abe:2008fi}
H.~Abe, T.~Kobayashi, and H.~Ohki, ``{Magnetized orbifold models},''
  \href{http://dx.doi.org/10.1088/1126-6708/2008/09/043}{{\em JHEP} {\bfseries
  0809} (2008) 043},
\href{http://arxiv.org/abs/0806.4748}{{\ttfamily arXiv:0806.4748 [hep-th]}}.
%%CITATION = ARXIV:0806.4748;%%.

\bibitem{Fujimoto:2013xha}
Y.~Fujimoto, T.~Kobayashi, T.~Miura, K.~Nishiwaki, and M.~Sakamoto, ``{Shifted
  orbifold models with magnetic flux},''
  \href{http://dx.doi.org/10.1103/PhysRevD.87.086001}{{\em Phys.Rev.}
  {\bfseries D87} (2013) 086001},
\href{http://arxiv.org/abs/1302.5768}{{\ttfamily arXiv:1302.5768 [hep-th]}}.
%%CITATION = ARXIV:1302.5768;%%.

\bibitem{Abe:2013bca}
T.-H. Abe, Y.~Fujimoto, T.~Kobayashi, T.~Miura, K.~Nishiwaki, {\em et~al.},
  ``{$Z_N$ twisted orbifold models with magnetic flux},''
  \href{http://dx.doi.org/10.1007/JHEP01(2014)065}{{\em JHEP} {\bfseries 1401}
  (2014) 065},
\href{http://arxiv.org/abs/1309.4925}{{\ttfamily arXiv:1309.4925 [hep-th]}}.
%%CITATION = ARXIV:1309.4925;%%.

\bibitem{Abe:2014noa}
T.-h. Abe, Y.~Fujimoto, T.~Kobayashi, T.~Miura, K.~Nishiwaki, and M.~Sakamoto,
  ``{Operator analysis of physical states on magnetized $T^{2}/Z_{N}$
  orbifolds},'' \href{http://dx.doi.org/10.1016/j.nuclphysb.2014.11.022}{{\em
  Nucl. Phys.} {\bfseries B890} (2014) 442--480},
\href{http://arxiv.org/abs/1409.5421}{{\ttfamily arXiv:1409.5421 [hep-th]}}.
%%CITATION = ARXIV:1409.5421;%%.

\bibitem{Buchmuller:2015eya}
W.~Buchmuller, M.~Dierigl, F.~Ruehle, and J.~Schweizer, ``{Chiral fermions and
  anomaly cancellation on orbifolds with Wilson lines and flux},''
  \href{http://dx.doi.org/10.1103/PhysRevD.92.105031}{{\em Phys. Rev.}
  {\bfseries D92} no.~10, (2015) 105031},
\href{http://arxiv.org/abs/1506.05771}{{\ttfamily arXiv:1506.05771 [hep-th]}}.
%%CITATION = ARXIV:1506.05771;%%.

\bibitem{Matsumoto:2016okl}
Y.~Matsumoto and Y.~Sakamura, ``{Yukawa couplings in 6D gauge-Higgs unification
  on $T^2/Z_N$ with magnetic fluxes},''
  \href{http://dx.doi.org/10.1093/ptep/ptw058}{{\em PTEP} {\bfseries 2016}
  no.~5, (2016) 053B06},
\href{http://arxiv.org/abs/1602.01994}{{\ttfamily arXiv:1602.01994 [hep-ph]}}.
%%CITATION = ARXIV:1602.01994;%%.

\bibitem{Kobayashi:2017dyu}
T.~Kobayashi and S.~Nagamoto, ``{Zero-modes on orbifolds : magnetized orbifold
  models by modular transformation},''
  \href{http://dx.doi.org/10.1103/PhysRevD.96.096011}{{\em Phys. Rev.}
  {\bfseries D96} no.~9, (2017) 096011},
\href{http://arxiv.org/abs/1709.09784}{{\ttfamily arXiv:1709.09784 [hep-th]}}.
%%CITATION = ARXIV:1709.09784;%%.

\bibitem{Abe:2012fj}
H.~Abe, T.~Kobayashi, H.~Ohki, A.~Oikawa, and K.~Sumita, ``{Phenomenological
  aspects of 10D SYM theory with magnetized extra dimensions},''
  \href{http://dx.doi.org/10.1016/j.nuclphysb.2013.01.014}{{\em Nucl.Phys.}
  {\bfseries B870} (2013) 30--54},
\href{http://arxiv.org/abs/1211.4317}{{\ttfamily arXiv:1211.4317 [hep-ph]}}.
%%CITATION = ARXIV:1211.4317;%%.

\bibitem{Abe:2016jsb}
H.~Abe, T.~Kobayashi, K.~Sumita, and Y.~Tatsuta, ``{Supersymmetric models on
  magnetized orbifolds with flux-induced Fayet-Iliopoulos terms},''
  \href{http://dx.doi.org/10.1103/PhysRevD.95.015005}{{\em Phys. Rev.}
  {\bfseries D95} no.~1, (2017) 015005},
\href{http://arxiv.org/abs/1610.07730}{{\ttfamily arXiv:1610.07730 [hep-ph]}}.
%%CITATION = ARXIV:1610.07730;%%.

\bibitem{Abe:2014vza}
H.~Abe, T.~Kobayashi, K.~Sumita, and Y.~Tatsuta, ``{Gaussian Froggatt-Nielsen
  mechanism on magnetized orbifolds},''
  \href{http://dx.doi.org/10.1103/PhysRevD.90.105006}{{\em Phys.Rev.}
  {\bfseries D90} no.~10, (2014) 105006},
\href{http://arxiv.org/abs/1405.5012}{{\ttfamily arXiv:1405.5012 [hep-ph]}}.
%%CITATION = ARXIV:1405.5012;%%.

\bibitem{Abe:2013bba}
H.~Abe, T.~Kobayashi, H.~Ohki, K.~Sumita, and Y.~Tatsuta, ``{Flavor landscape
  of 10D SYM theory with magnetized extra dimensions},''
  \href{http://dx.doi.org/10.1007/JHEP04(2014)007}{{\em JHEP} {\bfseries 1404}
  (2014) 007},
\href{http://arxiv.org/abs/1307.1831}{{\ttfamily arXiv:1307.1831 [hep-th]}}.
%%CITATION = ARXIV:1307.1831;%%.

\bibitem{Abe:2015yva}
T.-h. Abe, Y.~Fujimoto, T.~Kobayashi, T.~Miura, K.~Nishiwaki, M.~Sakamoto, and
  Y.~Tatsuta, ``{Classification of three-generation models on magnetized
  orbifolds},'' \href{http://dx.doi.org/10.1016/j.nuclphysb.2015.03.004}{{\em
  Nucl. Phys.} {\bfseries B894} (2015) 374--406},
\href{http://arxiv.org/abs/1501.02787}{{\ttfamily arXiv:1501.02787 [hep-ph]}}.
%%CITATION = ARXIV:1501.02787;%%.

\bibitem{Fujimoto:2016zjs}
Y.~Fujimoto, T.~Kobayashi, K.~Nishiwaki, M.~Sakamoto, and Y.~Tatsuta,
  ``{Comprehensive analysis of Yukawa hierarchies on $T^2/Z_N$ with magnetic
  fluxes},'' \href{http://dx.doi.org/10.1103/PhysRevD.94.035031}{{\em Phys.
  Rev.} {\bfseries D94} no.~3, (2016) 035031},
\href{http://arxiv.org/abs/1605.00140}{{\ttfamily arXiv:1605.00140 [hep-ph]}}.
%%CITATION = ARXIV:1605.00140;%%.

\bibitem{Buchmuller:2015jna}
W.~Buchmuller, M.~Dierigl, F.~Ruehle, and J.~Schweizer, ``{Split symmetries},''
  \href{http://dx.doi.org/10.1016/j.physletb.2015.09.069}{{\em Phys. Lett.}
  {\bfseries B750} (2015) 615--619},
\href{http://arxiv.org/abs/1507.06819}{{\ttfamily arXiv:1507.06819 [hep-th]}}.
%%CITATION = ARXIV:1507.06819;%%.

\bibitem{Buchmuller:2017vho}
W.~Buchmuller and J.~Schweizer, ``{Flavor mixings in flux compactifications},''
  \href{http://dx.doi.org/10.1103/PhysRevD.95.075024}{{\em Phys. Rev.}
  {\bfseries D95} no.~7, (2017) 075024},
\href{http://arxiv.org/abs/1701.06935}{{\ttfamily arXiv:1701.06935 [hep-ph]}}.
%%CITATION = ARXIV:1701.06935;%%.

\bibitem{Buchmuller:2017vut}
W.~Buchmuller and K.~M. Patel, ``{Flavor physics without flavor symmetries},''
  \href{http://dx.doi.org/10.1103/PhysRevD.97.075019}{{\em Phys. Rev.}
  {\bfseries D97} no.~7, (2018) 075019},
\href{http://arxiv.org/abs/1712.06862}{{\ttfamily arXiv:1712.06862 [hep-ph]}}.
%%CITATION = ARXIV:1712.06862;%%.

\bibitem{Ishida:2017avx}
M.~Ishida, K.~Nishiwaki, and Y.~Tatsuta, ``{Brane-localized masses in magnetic
  compactifications},''
  \href{http://dx.doi.org/10.1103/PhysRevD.95.095036}{{\em Phys. Rev.}
  {\bfseries D95} no.~9, (2017) 095036},
\href{http://arxiv.org/abs/1702.08226}{{\ttfamily arXiv:1702.08226 [hep-th]}}.
%%CITATION = ARXIV:1702.08226;%%.

\bibitem{Higaki:2016ydn}
T.~Higaki and Y.~Tatsuta, ``{Inflation from periodic extra dimensions},''
  \href{http://dx.doi.org/10.1088/1475-7516/2017/07/011}{{\em JCAP} {\bfseries
  1707} no.~07, (2017) 011},
\href{http://arxiv.org/abs/1611.00808}{{\ttfamily arXiv:1611.00808 [hep-th]}}.
%%CITATION = ARXIV:1611.00808;%%.

\bibitem{Buchmuller:2016dai}
W.~Buchmuller, M.~Dierigl, F.~Ruehle, and J.~Schweizer, ``{de Sitter vacua from
  an anomalous gauge symmetry},''
  \href{http://dx.doi.org/10.1103/PhysRevLett.116.221303}{{\em Phys. Rev.
  Lett.} {\bfseries 116} no.~22, (2016) 221303},
\href{http://arxiv.org/abs/1603.00654}{{\ttfamily arXiv:1603.00654 [hep-th]}}.
%%CITATION = ARXIV:1603.00654;%%.

\bibitem{Buchmuller:2016bgt}
W.~Buchmuller, M.~Dierigl, F.~Ruehle, and J.~Schweizer, ``{de Sitter vacua and
  supersymmetry breaking in six-dimensional flux compactifications},''
  \href{http://dx.doi.org/10.1103/PhysRevD.94.025025}{{\em Phys. Rev.}
  {\bfseries D94} no.~2, (2016) 025025},
\href{http://arxiv.org/abs/1606.05653}{{\ttfamily arXiv:1606.05653 [hep-th]}}.
%%CITATION = ARXIV:1606.05653;%%.

\bibitem{Ibanez:2006da}
L.~E. Ibanez and A.~M. Uranga, ``{Neutrino Majorana Masses from String Theory
  Instanton Effects},''
  \href{http://dx.doi.org/10.1088/1126-6708/2007/03/052}{{\em JHEP} {\bfseries
  03} (2007) 052},
\href{http://arxiv.org/abs/hep-th/0609213}{{\ttfamily arXiv:hep-th/0609213
  [hep-th]}}.
%%CITATION = HEP-TH/0609213;%%.

\bibitem{Ibanez:2012zz}
L.~E. Ibanez and A.~M. Uranga, {\em {String theory and particle physics: An
  introduction to string phenomenology}}.
\newblock Cambridge University Press, 2012.

\bibitem{Hamada:2014hpa}
Y.~Hamada, T.~Kobayashi, and S.~Uemura, ``{Flavor structure in D-brane models:
  Majorana neutrino masses},''
  \href{http://dx.doi.org/10.1007/JHEP05(2014)116}{{\em JHEP} {\bfseries 05}
  (2014) 116},
\href{http://arxiv.org/abs/1402.2052}{{\ttfamily arXiv:1402.2052 [hep-th]}}.
%%CITATION = ARXIV:1402.2052;%%.

\bibitem{Kobayashi:2015siy}
T.~Kobayashi, Y.~Tatsuta, and S.~Uemura, ``{Majorana neutrino mass structure
  induced by rigid instantons on toroidal orbifold},''
  \href{http://dx.doi.org/10.1103/PhysRevD.93.065029}{{\em Phys. Rev.}
  {\bfseries D93} no.~6, (2016) 065029},
\href{http://arxiv.org/abs/1511.09256}{{\ttfamily arXiv:1511.09256 [hep-ph]}}.
%%CITATION = ARXIV:1511.09256;%%.

\bibitem{Abe:2008sx}
H.~Abe, K.-S. Choi, T.~Kobayashi, and H.~Ohki, ``{Three generation magnetized
  orbifold models},''
  \href{http://dx.doi.org/10.1016/j.nuclphysb.2009.02.002}{{\em Nucl.Phys.}
  {\bfseries B814} (2009) 265--292},
\href{http://arxiv.org/abs/0812.3534}{{\ttfamily arXiv:0812.3534 [hep-th]}}.
%%CITATION = ARXIV:0812.3534;%%.

\bibitem{Frere:2010ah}
J.-M. Frere, M.~Libanov, and F.-S. Ling, ``{See-saw neutrino masses and large
  mixing angles in the vortex background on a sphere},''
  \href{http://dx.doi.org/10.1007/JHEP09(2010)081}{{\em JHEP} {\bfseries 09}
  (2010) 081},
\href{http://arxiv.org/abs/1006.5196}{{\ttfamily arXiv:1006.5196 [hep-ph]}}.
%%CITATION = ARXIV:1006.5196;%%.

\bibitem{Pilaftsis:1999jk}
A.~Pilaftsis, ``{Leptogenesis in theories with large extra dimensions},''
  \href{http://dx.doi.org/10.1103/PhysRevD.60.105023}{{\em Phys. Rev.}
  {\bfseries D60} (1999) 105023},
\href{http://arxiv.org/abs/hep-ph/9906265}{{\ttfamily arXiv:hep-ph/9906265
  [hep-ph]}}.
%%CITATION = HEP-PH/9906265;%%.

\bibitem{Dudas:2005vn}
E.~Dudas, C.~Grojean, and S.~K. Vempati, ``{Classical running of neutrino
  masses from six dimensions},''
\href{http://arxiv.org/abs/hep-ph/0511001}{{\ttfamily arXiv:hep-ph/0511001
  [hep-ph]}}.
%%CITATION = HEP-PH/0511001;%%.

\bibitem{Dobrescu:2001ae}
B.~A. Dobrescu and E.~Poppitz, ``{Number of fermion generations derived from
  anomaly cancellation},''
  \href{http://dx.doi.org/10.1103/PhysRevLett.87.031801}{{\em Phys. Rev. Lett.}
  {\bfseries 87} (2001) 031801},
\href{http://arxiv.org/abs/hep-ph/0102010}{{\ttfamily arXiv:hep-ph/0102010
  [hep-ph]}}.
%%CITATION = HEP-PH/0102010;%%.

\bibitem{Lim:2018lgg}
C.~S. Lim, ``{The implication of gauge-Higgs unification to the hierarchical
  fermion masses},''
\href{http://arxiv.org/abs/1801.01639}{{\ttfamily arXiv:1801.01639 [hep-ph]}}.
%%CITATION = ARXIV:1801.01639;%%.

\bibitem{Buchmuller:2016gib}
W.~Buchmuller, M.~Dierigl, E.~Dudas, and J.~Schweizer, ``{Effective field
  theory for magnetic compactifications},''
  \href{http://dx.doi.org/10.1007/JHEP04(2017)052}{{\em JHEP} {\bfseries 04}
  (2017) 052},
\href{http://arxiv.org/abs/1611.03798}{{\ttfamily arXiv:1611.03798 [hep-th]}}.
%%CITATION = ARXIV:1611.03798;%%.

\bibitem{Ghilencea:2017jmh}
D.~M. Ghilencea and H.~M. Lee, ``{Wilson lines and UV sensitivity in magnetic
  compactifications},'' \href{http://dx.doi.org/10.1007/JHEP06(2017)039}{{\em
  JHEP} {\bfseries 06} (2017) 039},
\href{http://arxiv.org/abs/1703.10418}{{\ttfamily arXiv:1703.10418 [hep-th]}}.
%%CITATION = ARXIV:1703.10418;%%.

\bibitem{Abe:2015uma}
H.~Abe, T.~Kobayashi, Y.~Tatsuta, and S.~Uemura, ``{D-brane instanton induced
  $\mu$ terms and their hierarchical structure},''
  \href{http://dx.doi.org/10.1103/PhysRevD.92.026001}{{\em Phys. Rev.}
  {\bfseries D92} no.~2, (2015) 026001},
\href{http://arxiv.org/abs/1502.03582}{{\ttfamily arXiv:1502.03582 [hep-ph]}}.
%%CITATION = ARXIV:1502.03582;%%.

\bibitem{Kobayashi:2016qag}
T.~Kobayashi, K.~Nishiwaki, and Y.~Tatsuta, ``{CP-violating phase on magnetized
  toroidal orbifolds},'' \href{http://dx.doi.org/10.1007/JHEP04(2017)080}{{\em
  JHEP} {\bfseries 04} (2017) 080},
\href{http://arxiv.org/abs/1609.08608}{{\ttfamily arXiv:1609.08608 [hep-th]}}.
%%CITATION = ARXIV:1609.08608;%%.

\bibitem{Esteban:2016qun}
I.~Esteban, M.~C. Gonzalez-Garcia, M.~Maltoni, I.~Martinez-Soler, and
  T.~Schwetz, ``{Updated fit to three neutrino mixing: exploring the
  accelerator-reactor complementarity},''
  \href{http://dx.doi.org/10.1007/JHEP01(2017)087}{{\em JHEP} {\bfseries 01}
  (2017) 087},
\href{http://arxiv.org/abs/1611.01514}{{\ttfamily arXiv:1611.01514 [hep-ph]}}.
%%CITATION = ARXIV:1611.01514;%%.

\bibitem{Capozzi:2017ipn}
F.~Capozzi, E.~Di~Valentino, E.~Lisi, A.~Marrone, A.~Melchiorri, and
  A.~Palazzo, ``{Global constraints on absolute neutrino masses and their
  ordering},'' \href{http://dx.doi.org/10.1103/PhysRevD.95.096014}{{\em Phys.
  Rev.} {\bfseries D95} no.~9, (2017) 096014},
\href{http://arxiv.org/abs/1703.04471}{{\ttfamily arXiv:1703.04471 [hep-ph]}}.
%%CITATION = ARXIV:1703.04471;%%.

\bibitem{Gariazzo:2018uwn}
S.~Gariazzo, M.~Archidiacono, P.~F. de~Salas, O.~Mena, C.~A. Ternes, and
  M.~T\'ortola, ``{Neutrino masses and their ordering: Global Data, Priors and
  Models},''
\href{http://arxiv.org/abs/1801.04946}{{\ttfamily arXiv:1801.04946 [hep-ph]}}.
%%CITATION = ARXIV:1801.04946;%%.

\bibitem{Lim:1990bp}
C.~S. Lim, ``{CP violation in higher dimensional theories},''
\href{http://dx.doi.org/10.1016/0370-2693(91)90679-K}{{\em Phys. Lett.}
  {\bfseries B256} (1991) 233--238}.
%%CITATION = PHLTA,B256,233;%%.

\bibitem{Lim:2009pj}
C.~S. Lim, N.~Maru, and K.~Nishiwaki, ``{CP Violation due to
  Compactification},'' \href{http://dx.doi.org/10.1103/PhysRevD.81.076006}{{\em
  Phys. Rev.} {\bfseries D81} (2010) 076006},
\href{http://arxiv.org/abs/0910.2314}{{\ttfamily arXiv:0910.2314 [hep-ph]}}.
%%CITATION = ARXIV:0910.2314;%%.

\bibitem{Fukugita:1986hr}
M.~Fukugita and T.~Yanagida, ``{Baryogenesis Without Grand Unification},''
\href{http://dx.doi.org/10.1016/0370-2693(86)91126-3}{{\em Phys. Lett.}
  {\bfseries B174} (1986) 45--47}.
%%CITATION = PHLTA,B174,45;%%.

\bibitem{Green:1984sg}
M.~B. Green and J.~H. Schwarz, ``{Anomaly Cancellation in Supersymmetric D=10
  Gauge Theory and Superstring Theory},''
\href{http://dx.doi.org/10.1016/0370-2693(84)91565-X}{{\em Phys. Lett.}
  {\bfseries 149B} (1984) 117--122}.
%%CITATION = PHLTA,149B,117;%%.

\end{thebibliography}\endgroup

\end{document}